\documentclass[letterpaper,11pt]{article}
\usepackage{jheppub}
\pdfoutput=1
\usepackage{empheq}

\usepackage{framed}
\definecolor{shadecolor}{rgb}{0.90,0.90,0.90}

\def\[{\left[}
\def\]{\right]}
\def\({\left(}
\def\){\right)}

\newcommand{\be}{\beta}

\def\Tr{\mathrm{Tr}}

\def\Tr{\mathrm{Tr}}

\def\tM{{{\mathtt M}}}
\def\tN{{{\mathtt N}}}

\def \be {\begin{equation}}
\def \ee {\end{equation}}
\def \bea {\begin{eqnarray}}
\def \eea {\end{eqnarray}}
\def \beal#1 {\begin{align}#1\end{align}}

\newcommand{\tsigma}{{\tilde \sigma}}
\newcommand{\ttau}{{\tilde \tau}}

\newcommand{\ZZ}{{\cal Z}}
\newcommand{\BB}{{\cal B}}
\newcommand{\os}{{\Omega}}
\newcommand{\oss}{{\tilde \Omega}}
\everymath{\displaystyle}

\preprint{TIFR/TH/20-13}

\title{Modularity of supersymmetric partition functions}

\author{Abhijit Gadde}
\affiliation{Department of Theoretical Physics,\\ Tata Institute of Fundamental Research, Mumbai 400005, India}
\emailAdd{abhijit@theory.tifr.res.in}

\abstract{
We discover a modular property of supersymmetric partition functions of supersymmetric theories with R-symmetry in four dimensions. This modular property is, in a sense, the generalization of the modular invariance of the supersymmetric partition function of two-dimensional supersymmetric theories on a torus i.e. of the elliptic genus. 
The partition functions in question are on manifolds homeomorphic to the ones obtained by gluing solid tori. Such gluing involves the choice of a large diffeomorphism of the boundary torus, along with the choice of a large gauge transformation for the background flavor symmetry connections, if present. 
Our modular property is a manifestation of the consistency of the gluing procedure. The modular property is used to rederive a supersymmetric Cardy formula for four dimensional gauge theories that has played a key  role  in computing the entropy of supersymmetric black holes. 
To be concrete, we work with four-dimensional ${\cal N}=1$ supersymmetric theories but we expect versions of our result to apply more widely to supersymmetric theories in other dimensions.
}

\begin{document}

\maketitle
\flushbottom

\section{Introduction}

Modular invariance of the torus partition function of two-dimensional theories has played an important role in constraining the spectrum of operators at high energy  \cite{CARDY1986186}. 
More recently, the program of modular bootstrap has been effective in obtaining other constraints on $2d$ conformal field theories such as on the gap between the vacuum and the first excited state. The modular property of the two-dimensional partition function on the torus ${\mathbb T}^2$ stems from ${\mathbb T}^2$ having a non-trivial large diffeomorphism group.  

It would be desirable to have a similar modular formula constraining the high energy spectrum of conformal field theories in higher dimensions. Such a formula has been lacking for a good reason. The spectrum of operators of a conformal field theory is captured by the thermal partition function i.e. the partition function on $S^{d-1}\times S^1$. For $d\neq 2$, this manifold does not have a non-trivial large diffeomorphism group. However, for some values of $d>2$, we can think of $S^{d-1}$ as a torus fibration. For $d=4$, it is ${\mathbb T}^2$ fibration over an interval and for $d=6$, it is ${\mathbb T}^3$ fibration over a disc\footnote{$S^5$ is an $S^1$ fibration over ${\mathbb {CP}}^2$ and ${\mathbb {CP}}^2$ is a toric manifold so it is a ${\mathbb T}^2$ fibration over a disc.}. In this paper, we will focus on the case of $d=4$ and briefly comment on the case of $d=6$. 
To be concrete, throughout the paper we will work with supersymmetric theories because supersymmetry allows an exact computation of relevant  partition functions. Our ideas, abstractly, seem applicable to even non-supersymmetric theories although concrete realizations in this case would be highly desirable. 

 In four dimensions, $S^3\times S^1$ is a  ${\mathbb T}^3$ fibration over an interval. This means it can be thought of as being obtained by gluing two solid three-tori ${\mathbb T}^2 \times {D}^2$. A more general gluing of the solid three-tori along the boundary ${\mathbb T}^3$ involves a choice of the large diffeomorphism of the boundary ${\mathbb T}^3$. This group is $SL(3,{\mathbb Z})$. If we pick the gluing element to be identity, we get $S^2 \times {\mathbb T}^2$. For a fancier choice of the gluing, we get lens spaces $L(r,s)\times S^1$. This set includes $S^3\times S^1$. In the paper, we will obtain a relation between the partition function obtained on the geometry obtained by gluing with $g_1g_2$ element with partition functions on geometries obtained by gluing with $g_1$ and $g_2$ separately. For a special choice of the group elements, this relation ends up being a modular property. For theories with global symmetry, we can turn on background global symmetry holonomies as allowed by supersymmetry. In that case, along with a choice of the large diffeomorphism, we also have to choose a large gauge transformation while gluing the two solid tori. In that case, $g_i\in {\cal G}$ where ${\cal G}$ is a group of large transformations: large diffeomorphisms times large gauge transformations.

In \cite{Festuccia:2011ws}, it was demonstrated how to preserve supersymmetry on a curved space by coupling the theory to supergravity and turning on the appropriate background. In \cite{Closset:2013vra}, the geometries on which two supercharges of ${\cal N}=1$ supersymmetric theory can be preserved were listed and studied. Happily, $S^2 \times {\mathbb T}^2$ and the lens spaces $L(r,s)$  do belong to this list. However, preserving supersymmetry on $S^2 \times {\mathbb T}^2$ involves turning on one unit of R-symmetry flux through $S^2$. This, in turn, means that the R-charges need to be quantized as integers. Moreover, to preserve supersymmetry on $L(r,s)$, the R-charges need to be quantized in units of $2/(s-1)$. If we insist on preserving supersymmetry on all lens spaces then we must require the R-charges to be even integers. If we require to preserve supersymmetry only under gluings by a certain subgroup of $\cal G$ which does not give rise to lens spaces then we can stick to R-charges only being integers. But note that the R-charges need to be quantized at least as integers to preserve supersymmetry under gluing because this is required by gluing with the identity element itself. It was also argued in \cite{Festuccia:2011ws} that the supersymmetric partition functions on manifolds that admit them are invariant under the RG flow. Hence the supersymmetric partition functions of a supersymmetric theory are, in fact, the partition functions of the endpoint of the RG flow i.e. of the superconformal field theory. 

The supersymmetric torus partition function in two dimensions is not completely modular invariant. It is modular invariant only up to a certain phase factor. This phase factor can not be removed by a local counterterm. It is the anomaly under large diffeomorphisms.   It is perhaps more apt to compare our setup with the supersymmetric partition function on ${\mathbb T}^2$ i.e. with the elliptic genus of, say, a $(2,2)$ supersymmetric theory. It is known to be a Jacobi form\footnote{If the target space of the $2d$ theory is not compact then there could be supersymmetric states that form a continuum. In such cases, the elliptic genus has a more complicated modular property. It is known to be a mock modular form.}. The phase factor depends only on the central charge and is known as the factor of automorphy. For a Jacobi form, the factor of automorphy is characterized only by two numbers, the weight (which is zero for elliptic genus) and the index (which is $c/6$ where $c$ is the central charge). The problem of computing the solution space of Jacobi forms with a given weight and index is a classic  problem and is a subject of an entire book \cite{eichler1985theory}.

In the same vein, the modular property that we find in four dimensions is also not exact\footnote{pun intended.} and does involve failure by phase factors coming from similar anomalies. 
The problem of computing the solution space of partition functions in four dimensions given the ``factor of automorphy" is a physically important one.  Because we expect that the factor of automorphy depends only on the anomalies of the theory, this problem would be tantamount to finding supersymmetric partition functions of theories with a given set of anomalies. 
If the supersymmetric partition functions - remember they are invariant under renormalization group flow -  
can be thought of as the proxy for the superconformal theories themselves then this problem becomes a classification program for four-dimensional superconformal theories. 

There has been a hint of a connection between anomalies, $SL(3,{\mathbb Z})$ and the superconformal index \cite{Spiridonov:2012ww}. In this paper, we uncover this connection and make it completely transparent.

\subsection*{Outline}
The rest of the paper is organized as follows. In section \ref{2d} we review modular properties of the two-dimensional supersymmetric partition functions with an emphasis on the group cohomological aspect. Then we move to the main part of the paper in section \ref{4d} and derive constraints on four-dimensional partition functions that follow from cutting and gluing. In doing this we think of the partition function on the compact manifold as the inner product of states on the torus boundaries of the two solid three-tori. After a quick discussion of the large diffeomorphism group of ${\mathbb T}^3$, $SL(3,{\mathbb Z})$, we will develop the idea for the chiral multiplet with R-charge $0$. While emphasizing the group cohomological aspect, we will make a connection with the so-called ``holomorphic block decomposition". In section \ref{examples} we will verify the modular properties for other theories. Examples consist of chiral multiplet with general R-charge (quantized appropriately according to the earlier discussion) and SQED with $N$ flavors. In section \ref{global-grav} we explain the relation between group cohomology and global gravitational and gauge anomalies. In section \ref{applications}, we give some applications of our modular formula. They include a bootstrap program for four dimensional superconformal field theories and a Cardy formula. We also describe the generalization of our four dimensional program to six dimensions. We end with some outlook.  
In the only appendix we list useful properties of some special functions.

\section{Two dimensions}\label{2d}
Before we get into the modular properties of the four-dimensional partition functions, let us review the modular properties of the ${\mathbb T}^2$ partition function of two-dimensional theories, i.e. of the elliptic genus. For concreteness, let us consider theories with ${\cal N}=(2,2)$ supersymmetry. The superconformal index is defined as
\be
{\cal Z}(z,\tau)={\Tr}_{\tt RR} \, (-1)^F e^{2\pi i z J} e^{2\pi i \tau L_0}.
\ee
Here $J$ is the charge of operators under the left-moving R-symmetry and $L_0$ is the left-moving conformal dimension. We can turn on additional background holonomies for global symmetries if they are present. For now, let us just assume that we have turned on holonomy only for $J$.
 The trace is taken over the ${\tt RR}$ sector. From path integral point of view, this index is the twisted partition function of the theory on a torus with periodic boundary conditions along both cycles. This supersymmetric partition function enjoys invariance under large diffeomorphisms $SL(2,{\mathbb Z})$ as well as under large gauge transformations ${\mathbb Z}^2$. The group of large symmetries forms the semi-direct product $SL(2,{\mathbb Z})\ltimes {\mathbb Z}^2\equiv {\cal G}^{(2d)}$. Its action on the parameters $(z,\tau)$ is
\bea\label{G-action}
&&g_1\cdot (z,\tau)= (\frac{z}{c\tau+d},\frac{a\tau+b}{c\tau+d}),\qquad\, g_1=
\begin{pmatrix}
a \,\,& b\\
c \,\, &d
\end{pmatrix}
\in SL(2,{\mathbb Z}),\nonumber \\
&&g_2\cdot (z,\tau)=(z+n_1+n_2\tau,\tau),\qquad g_2=(n_1,n_2)\in {\mathbb Z}^2.
\eea

In case there are additional background holonomies for global symmetries, then each will contribute ${\mathbb Z}^2$ to the group of large gauge transformations corresponding to shifts by $1$ and $\tau$ respectively. The group of large symmetries with $r$ fugacities turned on is $SL(2,{\mathbb Z})\ltimes ({\mathbb Z}^2)^r$. We stick with $r=1$. We assume that the target space of the theory is compact. For such theories, the superconformal index is a meromorphic function of $(z,\tau)$. For future convenience, let us denote the space of meromorphic functions by $\tN$. This space is endowed with the action \eqref{G-action} of ${\cal G}^{(2d)}$ on $(z,\tau)$  which makes $\tN$ a module of ${\cal G}^{(2d)}$. To be concrete we define the module action as $g\cdot \hat \ZZ(z,\tau)\equiv \hat \ZZ(g^{-1}\cdot(z,\tau))$.
The index is not exactly invariant but is invariant only up to a phase that captures anomalies of the theory.
\begin{shaded}
\be\label{genus}
{\cal Z}(z,\tau)= e^{i\phi_g(z,\tau)}\, {\cal Z}(g^{-1}\cdot(z,\tau)),\qquad g\in {\cal G}^{(2d)}.
\ee
\end{shaded} 
\noindent 
where $\phi_g(z,\tau)$ is a phase that encodes the anomalies of theory under ${\cal G}^{(2d)}$. As we are dealing with meromorphic functions of $(z,\tau)$, it is convenient to think of $e^{i\phi_g(z,\tau)}$ as an element of  the space $\tM$ of nowhere vanishing holomorphic functions. Just like $\tN$, the space $\tM$ is also a ${\cal G}^{(2d)}$ module. The module action is $g\cdot \phi_{g'}(z,\tau)\equiv \phi_{g'}(g^{-1}\cdot(z,\tau))$.
For consistency of equation \eqref{genus} $\phi_g(z,\tau)$ has to obey the group 1-cocycle condition:
\begin{shaded}
\be\label{sl2-phase}
e^{i\phi_{g_1\cdot g_2}(z,\tau)}= e^{i\phi_{g_1}(z,\tau)} \, e^{i\phi_{g_2}(g_1^{-1}\cdot(z,\tau))}.
\ee
\end{shaded}
\noindent As a result the values of $\phi_g$ are completely fixed by its values on the group generators. We have \cite{Kawai:1993jk},
\be
\phi_S(z,\tau)=2\pi i\frac{c}{6}\frac{z^2}{\tau} ,\qquad \phi_T(z,\tau)=0,\qquad \phi_{t_{1}}(z,\tau)=0.
\ee
Here $c$ is the central charge (either left-moving or right-moving). Elements $S$ and  $T$ are the generators of the modular group $SL(2,{\mathbb Z})$,  $S:(z,\tau) \to (z/\tau,-1/\tau)$ and  $T:(z,\tau)\to (z,\tau+1)$ and $t_1$ corresponds to the large gauge transformation $t_{1}:(z,\tau)\to (z+1,\tau)$. For future reference, let us also define the other large gauge transformation $t_\tau:(z,\tau)\to (z+\tau,\tau)$. Using the cocycle condition \eqref{sl2-phase}, we get
\be
\phi_{t_\tau}(z,\tau)=-2\pi i\frac{c}{6}(2z+\tau).
\ee

As the phase $\phi_g(z,\tau)$ encodes the anomalies under large transformations, it can not be removed by a local counter term. This means there does not exist a redefinition of the index  $\ZZ(z,\tau)$ which absorbs $\phi_g(z,\tau)$ for all group elements $g\in {\cal G}^{(2d)}$. One way to see this is as follows.
Consider $g_i=t_1^{m_i} t_\tau^{n_i}$ for $i=1,2$. Even though equation \eqref{sl2-phase} does hold,
\be\label{f-equation}
i\phi_{g_1g_2}(z,\tau)-i\phi_{g_1}(z,\tau)-i\phi_{g_2}(g_1^{-1}\cdot(z,\tau))=2\pi i m_1 n_2 \neq 0.
\ee
It is easy to see that the right hand side of equation \eqref{f-equation} would have been identically zero if we could absorb away all the phases into the partition function. This makes $e^{i\phi_g(z,\tau)}$, a nontrivial class in $H^1({\cal G}^{(2d)},\tM)$. 
The presence of the phase $e^{i\phi_g(z,\tau)}\in H^1({\cal G}^{(2d)},\tM)$ means that the index is not a function of the parameters $(z,\tau)$ but rather a section of a non-trivial bundle. It captures the anomaly of the theory under large diffeomorphisms and large gauge transformations.

The modules $\tN$ and $\tM$ form a short exact sequence, 
\be
1\to \tM\to \tN\to \tN/\tM\to 1.
\ee
It gives rise to the following long exact sequence for the cohomology groups,
\be
\ldots \to  H^j({\cal G}^{(2d)},\tM) \xrightarrow{i_*} H^j({\cal G}^{(2d)},\tN) \xrightarrow{p_*} H^j({\cal G}^{(2d)},\tN/\tM) \xrightarrow{\delta_*} H^{j+1}({\cal G}^{(2d)},\tM) \xrightarrow{i_*} \ldots. 
\ee
Here $\delta_*$ is the homomorphism defined by equation \eqref{genus}. It is then clear that if $e^{i\phi_g(z,\tau)}$ is nontrivial  in $H^1({\cal G}^{(2d)},\tM)$ then $\ZZ(z,\tau)$ is nontrivial in $H^0({\cal G}^{(2d)},\tN/\tM)$. The object $\ZZ(z,\tau)$ satisfying equation \eqref{genus} is also known in mathematics literature as the Jacobi form or more generally as ${\cal G}^{(2d)}$ automorphic form of degree 0 with ``the factor of automorphy" being $\phi_g(z,\tau)$. Given the factor of automorphy $e^{i\phi_g(z,\tau)}\in \tM$, the problem of finding the partition function $\ZZ(z,\tau)$ in $\tN/\tM$ is an interesting one. As remarked earlier, this problem is the subject of the book \cite{eichler1985theory}.

Global gravitational anomalies for fermion theories in $d$-dimensions have been discussed in a seminal paper by Witten \cite{Witten:1985xe}. These are classified by the so-called ``eta-invariant" \cite{aps73,aps75} on a $d+1$-dimensional manifolds. 
The case of the global gravitational anomaly for two-dimensional theories, in particular, has been studied in \cite{Witten:1985mj}. Recently, global gravitational anomalies were applied to classify symmetry protected topological phases \cite{Chen:2011pg}. In this context, group cohomology replaces the eta invariant in classifying anomalies in global transformations \cite{Witten:2019bou}. As exhibited by the above example, we find that group cohomology also plays an important role in classifying supersymmetric partition functions in two dimensions  through anomalies. We will find this to be the case even in four dimensions. 

With the brief discussion of the group cohomology above and illustration of its usefulness in classifying supersymmetric partition function in $2d$, we can spoil the punchline of the paper for the benefit of an eager and mathematically initiated reader: 
``The normalized part of the supersymmetric index" of a four-dimensional ${\cal N}=1$ supersymmetric field theory is a non-trivial class in  $H^1({\cal G},\tN/\tM)$. Here ${\cal G}$ is the group of large transformations i.e. large diffeomorphisms and large gauge transformations of ${\mathbb T}^3$ with background global symmetry holonomies. In the case of a theory with rank $r$ global symmetry, 
${\cal G}=SL(3,{\mathbb Z})\ltimes ({\mathbb Z}^3)^r$. What we mean by ``the normalized part of the supersymmetric index" will become clear in due course.

\section{Four dimensions}\label{4d}
Now we turn to the case of  four-dimensional ${\cal N}=1$ supersymmetric theory. We will assume that the theory has a single abelian global symmetry. The supersymmetric index is defined as \cite{Festuccia:2011ws}
\be\label{4dindex}
{\cal I}(z,\tau,\sigma)={\rm Tr}\, (-1)^F z^{J} p^{h_1-\frac{R}{2}}q^{h_2-\frac{R}{2}},\qquad x=e^{2\pi i z},\,p=e^{2\pi i \sigma},\,q=e^{2\pi i \tau}.
\ee
Here, $J$ is the charge under the global symmetry, $R$ is the $U(1)$ R-charge and $(h_1,h_2)$ are the Cartan generators of $SO(4)$ rotational symmetry. When the supersymmetric theory also has conformal symmetry, the supersymmetric index \eqref{4dindex} becomes the superconformal index. 
However, it is evaluated with a ``non-standard" choice of R-symmetry because the R-symmetry with integer (or even-integer) charges will not, in general, coincide with the superconformal R-symmetry. In order to compute the  true superconformal index, the integral R-symmetry needs to be shifted appropriately by abelian global symmetry so that the Weyl anomaly coefficient $a$ is maximized. This is achieved by a shift of global symmetry holonomies.

From the path integral point of view, the index is the twisted partition function on $S^3\times S^1$ with periodic boundary conditions for fermions along the $S^1$. As remarked earlier, this geometry does not have any interesting large diffeomorphisms. On the other hand it is observed that the index of a chiral multiplet (with R-charge $0$) is the elliptic gamma function $\Gamma(z,\tau,\sigma)$. This function is defined and its properties  listed in appendix \ref{functions}. In particular, it has an interesting modular property,
\begin{shaded}
\be\label{gamma-modular}
\Gamma(z,\tau,\sigma)\Gamma(\frac{z}{\sigma},\frac{\tau}{\sigma},\frac{1}{\sigma})\Gamma(\frac{z}{\tau},\frac{1}{\tau},\frac{\sigma}{\tau})= e^{-i\frac{\pi}{3} {Q}(z,\tau,\sigma)}.
\ee
\end{shaded}
\noindent
where ${Q}(z,\tau,\sigma)$ is a cubic polynomial in $z$ given in equation \eqref{sym-modular}.
Where is this modular property coming from? In \cite{Felder_2000}, the mathematical significance of this relation is explained. This property stems from $\Gamma(z,\tau,\sigma)$ being an automorphic form of degree $1$ of $SL(3,{\mathbb Z})\ltimes {\mathbb Z}^3$. 
To understand its physical origin, we have to think of the $S^3\times S^1$ geometry as being obtained by gluing together two solid ${\mathbb T}^3$. We give the main idea below.

\subsection{Main idea}
This gluing follows the more familiar construction of  $S^3$ by gluing two solid ${\mathbb T}^2$ where the contractible cycle of one side is identified with the non-contractible cycle of the other side and vice versa. The trasnverse $S^1$ plays the role of a spectator.

We formalize the construction as follows. The supersymmetry preserving solid ${\mathbb T}^3$ geometry is parametrized by two complex parameters $(\tau,\sigma)$. There is also a complex background holonomy $z_i$ for each Cartan generator of the global symmetry. Let the number of $z_i$'s be $r$.
We will often denote the set of these parameters $(z_i,\tau,\sigma)$ collectively as $\vec\tau$. 
 The background involves a partial topological twist on the contractible disc which effectively renders the Hilbert space on the boundary ${\mathbb T}^3$ finite-dimensional. Let $n$ be the dimension of this Hilbert space.
There are multiple choices for the basis vectors. For example,  we can take different types of supersymmetric surface operators inserted at the core of the solid torus and the resulting states would span the effective Hilbert space or we can take the Higgs branch vacua of the massive theory to span the effective Hilbert space. The latter choice will turn out to be the most convenient. Let us label these states as $|i;\vec\tau\rangle$, where $i$ is an enumeration label.

\begin{figure}[h]
    \centering
    {\includegraphics[width=0.35\textwidth]{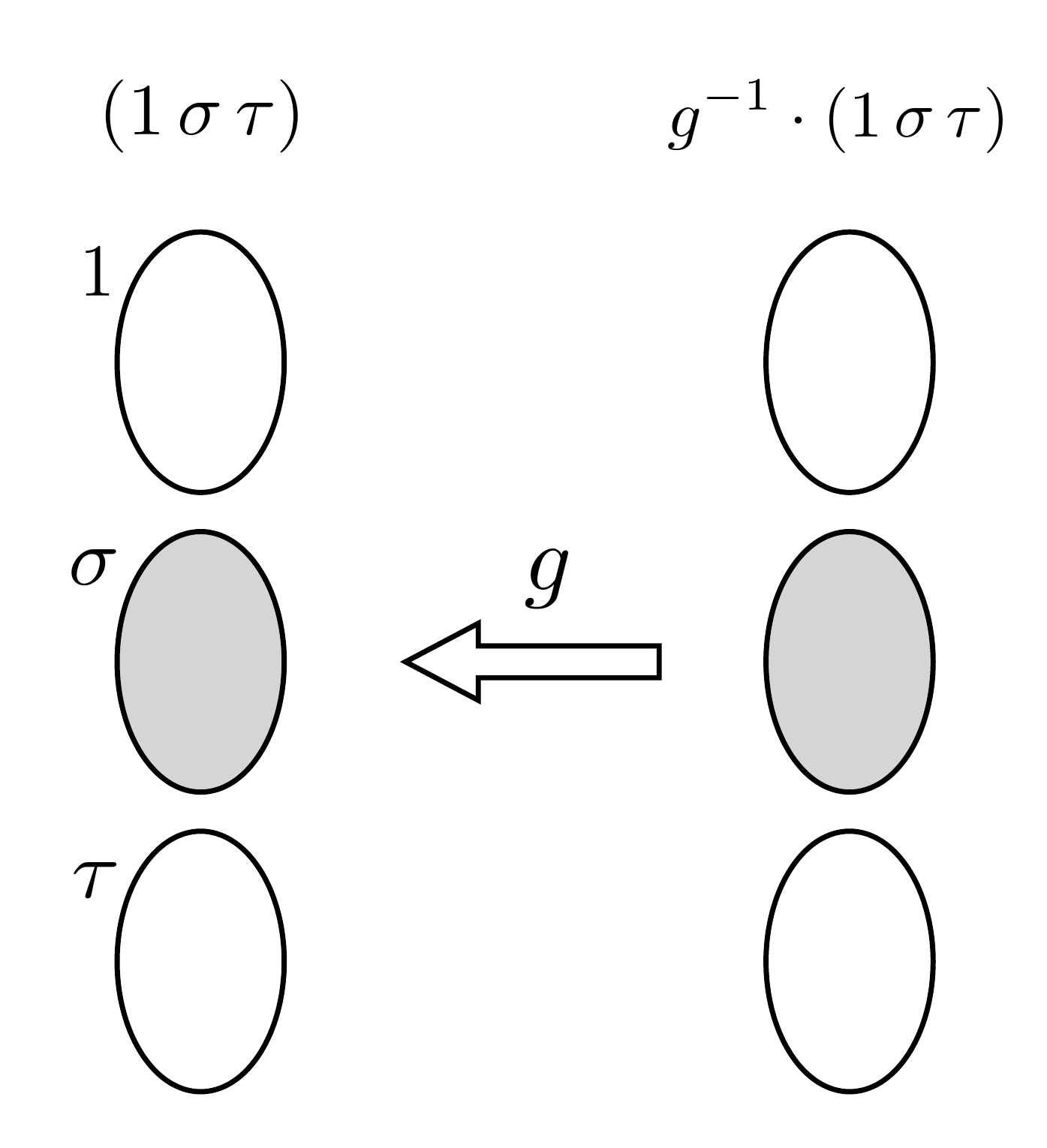} }
    \caption{Gluing of two solid three tori. We have shown the three cycles of each three-tori independently. The geometry is the direct product of these pieces. The cycle with length $\sigma$ that is contractible is shaded.}\label{gluing}
\end{figure}

The boundary torus has the large diffeomorphism group $SL(3,{\mathbb Z})$, it acts projectively as a $3\times 3$ matrix on column vector $(1\, \sigma\, \tau)^{T}$. The large gauge transformation group is $({\mathbb Z}^3)^r$, it is generated by the shift of $z_i, i=1,\ldots,r$ by $1,\sigma$ and $\tau$. Together they generate the semi-direct product ${\cal G}=SL(3,{\mathbb Z})\ltimes ({\mathbb Z}^3)^r$. Two solid three-tori can be glued by sandwiching $g\in {\cal G}$ to produce a compact geometry and a gauge bundle that preserves supersymmetry\footnote{The supersymmetry is preserved by making the anti-topological twist on the other half. This geometry is similar to the one considered in \cite{cecotti}. Its three-dimensional variation was used in \cite{Beem:2012mb}. We will not discuss this twist in detail.}. The gluing is schematically depicted in figure \ref{gluing}.
When this group element is chosen to be the $S$ generator of $SL(2,{\mathbb Z})\in SL(3,{\mathbb Z})$ acting on $(\sigma\, \tau)^T$, the resulting geometry is $S^3\times S^1$ and the partition function is the supersymmetric index.
The partition function on the more general background corresponding to gluing by $g$ is obtained by taking the inner product
\be
\ZZ_{g}(\vec\tau)=\langle \varnothing;\vec\tau|\varnothing;g^{-1}\cdot\vec\tau\rangle.
\ee
where $|\varnothing;\vec\tau\rangle$ is the vacuum state, corresponding to the solid three-torus with parameters $\vec\tau$ that is empty.
A more general observable for this background is the inner product $\langle i;\vec\tau|j;g^{-1}\cdot\vec\tau\rangle$. This corresponds to different choice of states for the two halves of the geometry. Let us call this the partition function matrix $\ZZ_{g}^{ij}$.
\be
\ZZ_{g}^{ij}(\vec\tau)\equiv \langle i;\vec\tau|j;g^{-1}\cdot\vec\tau\rangle.
\ee
We can make this object tractable using a complete orthonormal set of states $|\alpha\rangle$ that correspond to the exact supersymmetric ground states\footnote{of the supersymmetric quantum mechanics along the radial direction  obtained after the topological twist. See \cite{Beem:2012mb} for details in three dimensions.} on ${\mathbb T}^3$. Let us define,
\be
|\alpha; \vec\tau\rangle \equiv |\alpha\rangle \langle \alpha|\varnothing;\vec\tau\rangle\equiv |\alpha\rangle \BB_{\tt R}^{\alpha}(\vec \tau),\qquad {\rm also,}\quad \langle \varnothing;\vec\tau|\alpha\rangle\equiv \BB_{\tt L}^{\alpha}(\vec\tau).
\ee
These states correspond to Higgs branch vacua of the massive theory and $\BB_{\tt R}$, $\BB_{\tt L}$ are called the holomorphic blocks \cite{Peelaers:2014ima, yoshida2014factorization, Nieri:2015yia}. The partition function matrix is diagonal in this basis,
\be\label{factor}
\ZZ_{g}^{\alpha\beta} =\delta_{\alpha\beta} \,\BB_{\tt L}^{\alpha}(\vec\tau) \, \BB_{\tt R}^{\alpha}(g^{-1}\cdot \vec\tau).
\ee

Now we will obtain an equation relating the partition function matrices on different backgrounds obtained by gluing solid tori by large diffeomorphisms. 
\bea
\ZZ_{g_1g_2}^{\alpha\beta}(\vec \tau)&=& \langle \alpha; \vec \tau | \beta;g_2^{-1}g_1^{-1}\cdot \vec \tau\rangle\nonumber\\
&= & \sum_{\gamma=1}^{n} \sum_{\delta=1}^{n} \langle \alpha; \vec \tau | \gamma;g_1^{-1}\cdot \vec \tau\rangle  M_{\gamma \delta}     \langle \delta; g_1^{-1}\vec \tau     |\beta;g_2^{-1}g_1^{-1}\cdot \vec \tau\rangle\nonumber\\
&= & \sum_{\gamma=1}^{n} \sum_{\delta=1}^{n} \ZZ_{g_1}^{\alpha\gamma}(\vec \tau) M_{\gamma \delta} \ZZ_{g_2}^{\delta\beta}(g_1^{-1}\vec \tau).
\eea
In the second line we have inserted the complete set of states
\be
1=\sum_{\gamma=1}^{n} \sum_{\delta=1}^{n} | \gamma;g_1^{-1}\cdot \vec \tau\rangle   M_{\gamma \delta}    \langle \delta; g_1^{-1}\vec \tau  |.
\ee
The matrix $M$ is fixed by taking the matrix element of the identity with $\langle \alpha|1|\beta\rangle$.
\bea\label{step1}
\ZZ_1^{\alpha \beta}(g_1^{-1}\cdot \tau)&=& \ZZ_1^{\alpha \gamma}(g_1^{-1}\cdot \tau) M_{\gamma\delta} \ZZ_1^{\delta\beta}(g_1^{-1}\cdot \tau) \nonumber\\
\Rightarrow M_{\gamma \delta}&=& (\ZZ_1^{\gamma\delta}(g_1^{-1}\cdot \tau))^{-1}.
\eea
Define the partition function matrix with one upper and one lower index as 
\be
{\hat \ZZ}_g(\vec\tau) \equiv {\hat \ZZ}_{g,\beta}^{\alpha}(\vec \tau)\equiv  \ZZ_{g}^{\alpha\gamma}(\vec \tau) (\ZZ_1^{\gamma \beta}(g^{-1}\cdot \vec\tau))^{-1}.
\ee
The equation \eqref{step1} takes the form a matrix product,
\be\label{pre-master}
\hat \ZZ_{g_1g_2}(\vec \tau) =  \hat \ZZ_{g_1}(\vec \tau) \cdot \hat \ZZ_{g_2} (g_1^{-1}\cdot \vec\tau).
\ee
Because all the partition function matrices entering in the above discussion are diagonal in $|\alpha;\vec\tau\rangle$ basis. This equation is valid for all the entries in the diagonal separately.
\be\label{product}
\hat \ZZ_{g_1g_2}^{\alpha}(\vec \tau) =  \hat \ZZ_{g_1}^\alpha(\vec \tau) \cdot \hat \ZZ_{g_2}^\alpha (g_1^{-1}\cdot \vec\tau),\qquad \hat \ZZ_g^\alpha(\vec \tau)\equiv \ZZ_{g}^{\alpha\alpha}(\vec\tau)/\ZZ_1^{\alpha\alpha}(g^{-1}\cdot \vec\tau).
\ee
Alternatively, this equation can also be seen as a consequence of the factorization \eqref{factor} as this equation implies,
\be\label{trivial}
\hat \ZZ_{g}^{\alpha}(\vec\tau)=\BB^\alpha_{\tt L}(\vec\tau)/\BB^\alpha_{\tt L}(g^{-1}\cdot \vec\tau)
\ee
and equation \eqref{product} follows. As we will see shortly, this viewpoint is misleading.

In deriving equation \eqref{product}, 
we have assumed that the state $|\alpha,g\cdot {\vec \tau}\rangle$ depends on $g$ in a unique way. This assumption is not true if there is a non-trivial Berry connection on the space of parameter ${\vec \tau}$. As a result, after doing a loop in the parameter space to go to its $g$ image $|\alpha,\tau\rangle \to |\alpha,g\cdot {\vec \tau}\rangle$, the state will get multiplied by a phase. Moreover, this phase will depend on the path in the parameter space  if the Berry connection has curvature. In general, we should allow for such a possibility. This introduces a phase factor in equation \eqref{product}.
\begin{shaded}
\be\label{master}
\hat \ZZ_{g_1g_2}^\alpha(\vec \tau) =  e^{i\phi_{g_1,g_2}^\alpha(\vec \tau)}\hat \ZZ_{g_1}^\alpha(\vec \tau) \cdot \hat \ZZ_{g_2}^\alpha(g_1^{-1}\cdot \vec\tau).
\ee
\end{shaded}
\noindent The first thing to note is that, thanks to equation \eqref{master}, the phase satisfies the group cocycle condition 
\begin{shaded}
\be
e^{i\phi^\alpha_{g_1g_2,g_3}(\vec \tau)}=e^{i\phi_{g_1,g_2}^\alpha(\vec \tau)}e^{i\phi_{g_1,g_2g_3}^\alpha(\vec \tau)}e^{i\phi_{g_2,g_3}^\alpha(g_1^{-1}\cdot \vec \tau)}.
\ee
\end{shaded}
\noindent Moreover, if we multiply $\hat \ZZ_{g}^\alpha(\vec\tau)$ by $e^{if_g^\alpha(\vec\tau)}$ then the phase in equation \eqref{master} changes as,
\be
e^{i\phi_{g_1,g_2}^\alpha(\vec \tau)}\to e^{i\phi_{g_1,g_2}^\alpha(\vec \tau)}e^{i(f_{g_1 g_2}^\alpha(\vec\tau)-f_{g_1}^\alpha(\vec\tau)-f_{g_2}^\alpha(g_1^{-1}\cdot \vec\tau))}.
\ee 
If the Berry connection is curved then it is impossible to get rid of the phase in equation \eqref{master} with a simple redefinition of the partition function i.e. with a local counter terms as the Berry curvature is a sign of an anomaly. This means $e^{i\phi_{g_1,g_2}^\alpha(\vec \tau)}$ is a non-trivial element of $H^2({\cal G}, \tM)$. 
This makes $\hat \ZZ_g^\alpha(\vec\tau)$ a non-trivial element of $H^1({\cal G},\tN/\tM)$. The  object $\hat \ZZ_g^\alpha(\vec\tau)$ is also known as ${\cal G}$ automorphic form of degree $1$ with the factor of automorphy being $e^{i\phi_{g_1,g_2}^\alpha(\vec \tau)}$ (for all values of $g_1,g_2$).
The equation \eqref{factor} provides only a ``local trivialization" of the this cohomologically non-trivial element. Because \eqref{factor} is not valid ``globally", we expect it to fail for some generators of ${\cal G}$. We will see this explicitly in section \ref{free-chiral} and \ref{examples}.

Thanks to the diagonal nature of the partition function matrix in the $|\alpha;\vec\tau\rangle$ basis, we could  obtain the equation \eqref{master} for individual entries on the diagonal. The supersymmetric partition function is the trace of this matrix. In this sense, we think of each of the diagonal entry $\ZZ^{\alpha\alpha}_g(\vec\tau)$ of the partition function matrix as ``a part of the supersymmetric partition function". The partition function part that is appropriately normalized is then $\hat \ZZ^\alpha_g(\vec\tau)$.
This clarifies the phase ``a normalized part of the superconformal index" used at the end of section  \ref{2d}.

The phase $e^{i\phi_{g_1,g_2}^\alpha(\vec\tau)}$ appearing in the equation \eqref{master} is a type of global gravitational anomaly. It is more general than the type of gravitational anomaly discussed by Witten in \cite{Witten:1985xe}. It measures the obstruction to the consistency under cutting and gluing manifolds. It belongs to a class of anomalies that are sometimes called Dai-Freed anomalies. Such anomalies have appeared in the context of classifying symmetry protected topological phases in \cite{Chen:2011pg}. As remarked in \cite{Witten:2019bou}, group cohomology classifies the anomalies for global transformations for bosonic phases the way eta invariant classifies global anomalies for fermionic phases.
Although our analysis has allowed this phase to depend on the state $|\alpha\rangle$, we conjecture that it is independent of it. This is because we do not expect anomalies to depend on the state.

Just like the classification problem for the supersymmetric torus partition function in two dimensions, we envision a classification problem for four-dimensional (normalized) supersymmetric partition functions $\hat \ZZ_g(\vec\tau)$ given the factors of automorphy $\phi_{g_1,g_2}(\vec\tau)$. To our knowledge, unlike in the two-dimensional case, a systematic treatment of this problem is lacking. We believe it is very much wanting. We conjecture that the factors of automorphy depend only on the 't Hooft anomaly polynomial and give an explicit expression for it for a choice of the pair $(g_1,g_2)$. 

Now to make connection with the modular property of the elliptic Gamma function, we observe the following. Evaluating the equation \eqref{master} on the group relation $Y^3=1$, where $Y\in SL(3,{\mathbb Z})$ is an element which cyclically permutes $(1 \,\sigma\, \tau)$, gives 
\begin{shaded}
\bea\label{modularY}
{\hat \ZZ}_Y^\alpha(\vec \tau) {\hat \ZZ}_Y^\alpha(Y^{-1}\cdot\vec\tau) {\hat \ZZ}_Y^\alpha(Y^{-2}\cdot \vec\tau)&=&1\quad({\rm mod}\,\,\tM),\nonumber\\
&\equiv & e^{-i\frac{\pi}{3}P(\vec\tau)}.
\eea
\end{shaded}
\noindent 
As $\hat \ZZ_g^\alpha$ are elements of cohomology that are defined only modulo $\tM$, the second equality   is meaningful only if this multiplicative freedom is fixed. For an abelian gauge theory, we will show that $\hat \ZZ_Y^\alpha=Z_P^\alpha\,\,({\rm mod}\,\,\tM)$ in cohomology where $Z_P^\alpha$ is the perturbative part (defined in section \ref{examples}) of the supersymmetric index $\ZZ_{S_{23}}$. We expect this fact to be true for general gauge theories.
 For models involving only chiral multiplets, $Z_P^\alpha=\ZZ_{S_{23}}$. Then the representative that appears in equation \eqref{modularY} is taken to be $\hat \ZZ_Y^\alpha=Z_P^\alpha$ in cohomology.  This fixes the $({\rm mod}\,\,\tM)$ freedom and makes the second equality in equation \eqref{modularY} meaningful.
 In this way, the formula directly connects to a more physical observable.
We conjecture that the function $P(\vec \tau)$ is essentially the 't-Hooft anomaly polynomial of the theory. Concretely,
\begin{shaded}
\bea\label{anomalyY}
&&P\Big(\frac{\xi_i}{\omega_1},\frac{\omega_2}{\omega_1},\frac{\omega_3}{\omega_1}\Big)=\frac{1}{\omega_1\omega_2\omega_3}\Big( k_{ijk}\xi_i\xi_j\xi_k+3k_{ijR}\,\os \xi_i\xi_j+3 k_{iRR}\,\os^2 \xi_i\nonumber\\
&&+k_{RRR}\,\os^3-k_i\,\oss \xi_i-k_R \,\os\oss\Big);\quad
\os\equiv  \frac12\sum_{j=1,2,3}\omega_j,\,\, \oss \equiv  \frac14\sum_{j=1,2,3}\omega_j^2.
\eea
\end{shaded}
\noindent Here we have used homogeneous coordinates $(\xi_i,\omega_1,\omega_2,\omega_3)$  introduced shortly instead of the affine ones $(z_i,\sigma,\tau)$. The symbols $k_{ijk}={\rm Tr}(F_iF_jF_k), \,k_{ijR}={\rm Tr}(F_iF_jR)$ and so on and $k_i={\rm Tr}(F_i), \,k_R={\rm Tr}(R)$ are 't Hooft anomalies.
We will  offer substantial evidence in support of this conjecture. Equations \eqref{modularY}, \eqref{anomalyY} summarize  our proposal for the the modular property of the four-dimensional supersymmetric partition functions. 

For a chiral multiplet, the supersymmetric Hilbert space on ${\mathbb T}^3$ is one-dimensional and $\vec\tau=(z\,\sigma\,\tau)$. 
For the chiral multiplet with R-charge $0$, this equation precisely implies the modular property of the elliptic Gamma function \eqref{gamma-modular}. This is detailed in section \ref{free-chiral}. The above equations generalize this property to partition functions of general supersymmetric theories. 
Relatedly, an $SL(2,{\mathbb Z})$ modular property of the ``Schur-limit" of the ${\cal N}=2$ superconformal index has been discovered in \cite{Razamat:2012uv}.  It is also shown there that for a free Hypermultiplet, this property descends from the  $SL(3,{\mathbb Z})$ modular property \eqref{gamma-modular} of the elliptic gamma function. As equation \eqref{modularY} is the generalization of equation \eqref{gamma-modular} for interacting theories, we expect the $SL(2,{\mathbb Z})$ modular property of the Schur index of interacting theories to follow from equation \eqref{modularY} in the Schur limit. It would be interesting to solidify this connection.

Before moving to the demonstration of equation \eqref{master} for free and interacting theories, we recall certain basic facts about and set up the notation for $SL(3,{\mathbb Z})\ltimes {\mathbb Z}^3$.

\subsection{$SL(3,{\mathbb Z})\ltimes {\mathbb Z}^3$}

In this section, we describe the group of large diffeomorphisms and large gauge transformations for a theory with a single background holonomy turned on. Generalization to the case of multiple background holonomies is straightforward. The group of large symmetries of ${\mathbb T}^3$ with a single background holonomy is ${\cal G}_1=SL(3,{\mathbb Z})\ltimes {\mathbb Z}^3$. It is convenient to think of its action on a rectangular ${\mathbb T}^3$. Let the lengths of the three cycles be $(\omega_1\,\, \omega_2\,\, \omega_3)$.  We will think of this as a column vector on which elements of $SL(3,{\mathbb Z})$ act as $3\times 3$ matrix. Because we are only interested in projective representation of $SL(3,{\mathbb Z})$, physical observables only depend on $\omega_2/\omega_1\equiv \sigma, \omega_3/\omega_1\equiv \tau$. Using projective invariance we can scale $\omega_1=1$ and  think of  $(\omega_1\,\, \omega_2\,\, \omega_3)\simeq (1\,\,\sigma\,\, \tau)$. We will use projective coordinates $(\omega_1\,\, \omega_2\,\, \omega_3)$ and affine coordinates $(1\,\,\sigma\,\, \tau)$ interchangeably.
Remember that this ${\mathbb T}^3$ is actually the boundary of a solid three-torus. We take the contractible cycle to be the one with  length $\omega_2\simeq \sigma$. Let $z=\xi/\omega_1$ be the background holonomy.

It is convenient to identify and label the generators of ${\cal G}_1$. The standard choice of $SL(3,{\mathbb Z})$ generators consist of the three matrices $T_{ij},1\leq i\neq j\leq 3$ which have $1$ on the diagonal and at the $ij$-th place and $0$ everywhere else. For example,
\be
T_{12}=
\left(
\begin{array}{ccc}
1 & 1 & 0\\
0 & 1 & 0\\
0 & 0 & 1
\end{array}
\right),\qquad
T_{23}=
\left(
\begin{array}{ccc}
1 & 0 & 0\\
0 & 1 & 1\\
0 & 0 & 1
\end{array}
\right),\qquad {\rm etc.}
\ee
 In this presentation, $SL(3,{\mathbb Z})$ is generated by $T_{ij}$'s subjected to relations.
\be\label{standard}
T_{ij}T_{kl}=T_{kl}T_{ij} \quad(i\neq k,j\neq l),\quad T_{ij} T_{jk}=T_{ik} T_{jk} T_{ij},\quad (T_{13}T_{31}^{-1} T_{13})^4=1.
\ee

For our purposes, it would be useful to choose a different set of generators to make contact with the various $SL(2,{\mathbb Z}) $ subgroups. We define $S_{ij}, 0\leq i<j\leq 3$ to be the modular S-matrix for the $SL(2,{\mathbb Z})$ subgroup acting on the space $(ij)$. Explicitly,
\be
S_{12}=
\left(
\begin{array}{ccc}
0 & 1 & 0\\
-1 & 0 & 0\\
0 & 0 & 1
\end{array}
\right),\qquad 
S_{23}=
\left(
\begin{array}{ccc}
1 & 0 & 0\\
0 & 0 & 1\\
0 & -1 & 0
\end{array}
\right),\qquad
S_{13}=
\left(
\begin{array}{ccc}
0 & 0 & 1\\
0 & 1 & 0\\
-1 & 0 & 0
\end{array}
\right).
\ee
Then $SL(3,{\mathbb Z})$ is generated by $\{S_{23},S_{13},T_{23}\}$. It is easy to construct all the $T_{ij}$'s with these elements. First note that $S_{12}=S_{23}S_{13}S_{23}^{-1}$. Now all the other $T_{ij}$'s are constructed by  conjugating $T_{23}$ with all the $S_{ij}$'s. For example, $S_{13}T_{23}S_{13}^{-1}=T_{21}$ etc.. Once we have $T_{ij}$'s we can  generate the entire group thanks to the standard presentation \eqref{standard}. What are the relations? For each of the $SL(2,{\mathbb Z})$ subgroup, the $S$ and $T$ generators obey the usual relations,
\be\label{relations}
S_{ij}^4=1,\qquad (S_{ij} T_{ij})^3=1
\ee
In addition to these there is a relation that connects all the three subgroups. For this purpose we construct the element $Y\equiv S_{23}^{-1}S_{13}$. It permutes all the entries cyclically. It obeys the relation,
\be
Y^3=1.
\ee
This is  the group element that appears in the modular equation \eqref{modularY}.

The large gauge transformation subgroup ${\mathbb Z}^3$ is generated by shift operators $t_i, i=1,2,3$ which act on the holonomy $z$ as $z\to z+1$, $z\to z+\sigma$ and $z\to z+\tau$ respectively. 
After $SL(3,{\mathbb Z})$ is generated $SL(3,{\mathbb Z})\ltimes {\mathbb Z}^3$ is generated by adding a one of the $t_i$ to the list of $SL(3,{\mathbb Z})$ generators, say $t_3$. The other generators of ${\mathbb Z}^3$, $t_{1},\,t_{2}$, can be generated from $t_3$ by conjugating with $S_{ij}$'s.
 In conclusion, $SL(3,{\mathbb Z})\ltimes {\mathbb Z}^3$ is generated by $\{S_{23},S_{13},T_{23},t_3\}$.

We have described in detail the group of large symmetries ${\cal G}$ of ${\mathbb T}^3=\partial({\mathbb T}^2\times D^2)$. There exists a special subgroup $\cal H$ of transformations which can be extended into the bulk i.e. the subgroup of large symmetries of the solid three-torus ${\mathbb T}^2\times D^2$. We expect the wavefunction to be invariant under these large symmetries. Of course, this invariance is only up to a phase that captures the standard Witten type gravitational anomaly. It is easy to identify this subgroup. As this geometry consists of ${\mathbb T}^2$ spanned by cycles of length $(1\,\,\tau)$, we expect invariance under $SL(2,{\mathbb Z})$ acting  on $(1 \,\,\tau)^T$ subspace. We also expect invariance under the large gauge transformations $t_1,t_3$. Together this group is $SL(2,{\mathbb Z})\ltimes {\mathbb Z}^2$. In addition to these large symmetries, diffeomorphisms corresponding to $\sigma \to \sigma+1$ and $\sigma \to \sigma +\tau$ can also be extended in the bulk. These are the transformations $T_{21}$ and $T_{23}$ respectively. Together they   generate another factor of ${\mathbb Z}^2$ which is acted upon by $SL(2,{\mathbb Z})$. All in all, the subgroup $\cal H$ of large symmetries that can be extended into the bulk is $SL(2,{\mathbb Z})\ltimes ({\mathbb Z}^2)^2$. The $SL(2,{\mathbb Z})$ is  generated by $\{S_{13}, T_{13}\}$ in the standard way and the two  ${\mathbb Z}^2$s are generated by $\{t_{1},t_3\}$ and $\{T_{21}, T_{23}\}$ respectively.

Because, the wavefunction is unchanged (modulo $\tM$) under the action of $h\in {\cal H}$, we expect
\be\label{H-element}
\hat \ZZ_{h}^\alpha(\vec\tau)=1 \quad ({\rm mod}\,\,\tM),\qquad {\rm for}\quad h\in {\cal H}
\ee
Thanks to the the equation \eqref{master}, in order to compute ${\hat \ZZ}_g^\alpha$ for all $g\in {\cal G}$, we need to compute it only on the generators. Given $\hat \ZZ_h^\alpha(\vec \tau)=1$, we only need to compute it on $S_{23}$. In order to get this ``normalized" partition functions, we need to compute  the physical partition function $\ZZ_g^{\alpha\alpha}(\vec\tau)$ on $g\in \{1,  S_{23}\}$. Recall $\ZZ_{S_{23}}$ is the partition function  on $S^3\times S^1$ i.e. the superconformal index $\cal I $ and  $\ZZ_1$ is the partition function on $S^2\times {\mathbb T}^2$. 

The fact that we need partition functions $\ZZ_{S_{23}}$ and $\ZZ_{1}$ to compute partition functions over all lens spaces, has an interesting physical significance. The partition function $\ZZ_{S_{23}}$ of gauge theories depends only on the Lie algebra of the group group and is insensitive to its global properties. On the other hand,  the partition function on any other lens space, thanks to its non-trivial fundamental group, is sensitive to the global structure of the gauge group \cite{Razamat:2013opa}. In our formalism, this sensitivity comes from the partition function on $S^2\times {\mathbb T}^2$ i.e. $\ZZ_{1}$.

\section{A chiral multiplet in $4d$}\label{free-chiral}
For an ${\cal N}=1$ chiral multiplet with R-charge $0$, the effective Hilbert space if one-dimensional. Hence we will drop the superscript $\alpha$ on $\ZZ$ and $\hat \ZZ$. The required partition functions are,
\be\label{localization}
\ZZ_{1}(z,\sigma,\tau)=\frac{1}{\theta(z,\tau)},\qquad \ZZ_{S_{23}}(z,\sigma,\tau)=\Gamma(z,\sigma,\tau).
\ee
For $\ZZ_1$  we have borrowed the results from \cite{Closset:2013sxa}. From these physical  partition functions, we construct the normalized one, 
\be\label{generators}
\hat \ZZ_{S_{23}}(z,\sigma,\tau)=\Gamma(z+\tau,\sigma,\tau).
\ee
From \eqref{localization}, along with \eqref{H-element}, we can compute the partition function on any background obtained by gluing two copies of solid ${\mathbb T}^3$ by an element $g\in{\cal G}$. 
But before that let us first verify that these partition functions  are consistent with some of the group relations \eqref{relations}.
\bea
S_{23}^4=1\quad&\Rightarrow& \quad\Gamma(z+\tau,\sigma,\tau)\Gamma(z+\sigma,-\tau,\sigma)\Gamma(z-\tau,-\sigma,-\tau)\Gamma(z-\sigma,\tau,-\sigma)=1\quad ({\rm mod}\,\,\tM),\nonumber\\
S_{13}^4=1 \quad &\Rightarrow&\quad  1\cdot 1\cdot1\cdot 1=1\quad ({\rm mod}\,\,\tM),\nonumber\\
(S_{23}T_{23})^{3}=1\quad  &\Rightarrow& \quad \Gamma(z+\tau,\sigma,\tau)\Gamma(z+\sigma, -\sigma-\tau,\sigma)\Gamma(z,\tau,\sigma+\tau)=1\quad ({\rm mod}\,\,\tM).
\eea
In the second equation we have used $\hat \ZZ_{S_{13}}=1$ because $S_{13}\in \cal H$.
That the elliptic gamma function satisfies the above conditions can be checked easily from the properties listed in appendix \ref{elliptic-gamma}. 
In order to verify that the partition function respects the most non-trivial relation $Y^3=1$, we need to compute $\hat \ZZ_Y$.
\be\label{Y-S}
\hat \ZZ_Y(z,\sigma,\tau)=\hat \ZZ_{S_{23}^{-1}S_{13}}(z,\sigma,\tau)={\hat \ZZ}_{S_{23}^{-1}}(z,\sigma,\tau) \quad ({\rm mod}\,\,\tM).
\ee
Again we have used $\hat \ZZ_{S_{13}}=1$. The partition function ${\hat \ZZ}_{S_{23}^{-1}}$ is computed by using the relation $S_{23}^{-1}S_{23}=1$. It turns out,
\be
\hat \ZZ_Y(z,\sigma,\tau)={\hat \ZZ}_{S_{23}^{-1}}(z,\sigma,\tau)=\Gamma(z,\sigma,\tau)={ \ZZ}_{S_{23}}(z,\sigma,\tau) \quad ({\rm mod}\,\,\tM).
\ee
Taking $\hat \ZZ_Y(z,\sigma,\tau)={ \ZZ}_{S_{23}}(z,\sigma,\tau)$,  equation \eqref{modularY} implies,
\be\label{chiral0}
\quad \Gamma(z,\tau,\sigma)\Gamma(\frac{z}{\sigma},\frac{\tau}{\sigma},\frac{1}{\sigma})\Gamma(\frac{z}{\tau},\frac{1}{\tau},\frac{\sigma}{\tau})= e^{-i\frac{\pi}{3} Q(z,\sigma,\tau)}.
\ee
This explains the mysterious modular property of the elliptic gamma function \eqref{gamma-modular}.  The polynomial $Q(z,\sigma,\tau)$ is  given in equation \eqref{sym-modular}. It is precisely the anomaly polynomial $P(z,\sigma,\tau)$ for the chiral multiplet with R-charge $0$. This is shown explicitly near equation \eqref{phase-anomaly}.

We can compute partition functions on general manifolds obtained by gluing two solid tori by large diffeomorphism and large gauge transformation such as lens spaces. Let us see a simple example first. The element $t_2$ can be expressed as $t_2=S_{23}\,t_3\,S_{23}^{-1}$. From here, using equation \eqref{master}, we can compute the partition function $\ZZ_{t_2}$.
\bea
\hat \ZZ_{t_2}(z,\tau,\sigma) \hat \ZZ_{S_{23}}(t_2^{-1}\cdot(z,\tau,\sigma))&=&\hat \ZZ_{S_{23}}(z,\tau,\sigma) \hat \ZZ_{t_3}(S_{23}^{-1}(z,\tau,\sigma)) \nonumber\\
\hat \ZZ_{t_2}(z,\tau,\sigma)&=&\Gamma(z+\tau,\tau,\sigma)/\Gamma(z+\tau-\sigma,\tau,\sigma)\nonumber\\
&=&\theta(z+\tau-\sigma,\tau)=\theta(z-\sigma, \tau)\quad ({\rm mod}\,\,\tM)\nonumber\\
 \ZZ_{t_2}(z,\tau,\sigma)&=&1.
\eea
In the second line we have used $\hat \ZZ_{t_3}=1$. Is the conclusion $\ZZ_{t_2}(z,\tau,\sigma)=1$ correct? To answer this we first need to understand the effect of the large gauge transformation $t_2$ in gluing. As the $SL(3,{\mathbb Z})$ part of the gluing is trivial, the geometry is $S^2\times {\mathbb T}^2$. The large gauge transformation $t_2$ used for gluing produces one unit of magnetic flux through $S^2$ because this is precisely how one constructs the nontrivial $U(1)$ bundle over $S^2$, namely, gluing the locally trivial bundles on the two discs by large gauge transformation. Comparing with the result in \cite{Closset:2013sxa} for the chiral multiplet partition function on  $S^2\times {\mathbb T}^2$ with a single unit of magnetic flux, we see that our conclusion indeed agrees with it.

\subsection{Lens space index}\label{lens}
Now that we have verified that the normalized partition functions for group generators indeed satisfy (some of) the group relations, we will now go ahead and construct partition functions on other geometries obtained by $g$-gluing using equation \eqref{master}. All these computations will be valid modulo multiplication by a phase i.e. mod $\tM$. 

The superconformal index on $L(r,1)\times S^1$ has been studied in \cite{Benini:2011nc, Razamat:2013opa, Razamat:2013jxa, Kels:2017toi}. In this subsection, we will focus our attention to these geometries. The lens space $L(r,1)$ is obtained by gluing two solid 2-tori with the element  $g_r=ST^{-r}S$ of $SL(2,{\mathbb Z})$. This $SL(2,{\mathbb Z})$ can be thought of as the $SL(3,{\mathbb Z})$ subgroup that is either acting on $(1\,\,\sigma)^T$ subspace of $(\sigma\,\,\tau)^T$ subspace. We will take it to be the latter i.e. $g_r=S_{23}T_{23}^{-r} S_{23}$.  
\bea
{\hat \ZZ}_{S_{23}T_{23}^{-r} S_{23}}(z,\sigma,\tau)&=&{\hat \ZZ}_{S_{23}}(z,\sigma,\tau) {\hat \ZZ}_{T_{23}^{-r}}(S_{23}^{-1}\cdot (z,\tau,\sigma)){\hat \ZZ}_{S_{23}}(T_{23}^{r}S_{23}^{-1}\cdot(z,\sigma,\tau)).\nonumber\\
&=&{\hat \ZZ}_{S_{23}}(z,\sigma,\tau) {\hat \ZZ}_{S_{23}}(T_{23}^{r}S_{23}^{-1}\cdot(z,\sigma,\tau)).\nonumber\\
\frac{{\ZZ}_{S_{23}T_{23}^{-r} S_{23}}(z,\sigma,\tau)}{{\ZZ}_{1}(S_{23}^{-1}T_{23}^{r}S_{23}^{-1}\cdot(z,\sigma,\tau))}&=&\frac{{\hat \ZZ}_{S_{23}}(z,\sigma,\tau) {\ZZ}_{S_{23}}(T_{23}^{r}S_{23}^{-1}\cdot(z,\sigma,\tau))}{{\ZZ}_{1}(S_{23}^{-1}T_{23}^{r}S_{23}^{-1}\cdot(z,\sigma,\tau))}\nonumber\\
{\ZZ}_{S_{23}T_{23}^{-r} S_{23}}(z,\sigma,\tau)&=& {\hat \ZZ}_{S_{23}}(z,\sigma,\tau) {\ZZ}_{S_{23}}(T_{23}^{r}S_{23}^{-1}\cdot(z,\sigma,\tau))\nonumber\\
&=&\Gamma(z+\tau,\sigma,\tau)\Gamma(z,r\sigma-\tau,\sigma).
\eea
In the second line we have used ${\hat \ZZ}_{T_{23}}(z,\tau,\sigma)=1$ mod $\tM$ and in the third line we have changed from normalized partition function to the physical partition function. Lens space index for the chiral multiplet has been first computed in \cite{Benini:2011nc}. In order to match our expression with the expression there, we need to substitute,  $(\sigma,\tau)= (\tsigma+\ttau, r\ttau)$. Under this change of variables,
\be
\ZZ_{g_r}(z,\sigma,\tau)= \Gamma(z+r\ttau,\tsigma+\ttau ,r\ttau)\Gamma(z,\tsigma+\ttau,r\tsigma).
\ee

The Lens space index with $h$ units of background magnetic flux is also constructed straighforwardly by computing $\ZZ_{g_{r,h}}(z,\sigma,\tau)$ where, $g_{r,h}=t_2^h\,S_{23}T_{23}^{-r} S_{23}$. 
\bea
{\hat \ZZ}_{g_{r,h}}(z,\sigma,\tau)&=&{\hat \ZZ}_{t_2^h}(z,\sigma,\tau){\hat \ZZ}_{S}(t_2^{-h}\cdot(z,\sigma,\tau)) {\hat \ZZ}_S(T^{r}S^{-1}t_2^{-h}\cdot(z,\sigma,\tau)),\nonumber\\
{\ZZ}_{g_{r,h}}(z,\sigma,\tau)&=&{\hat \ZZ}_{t_2^h}(z,\sigma,\tau){\hat \ZZ}_{S}(t_2^{-h}\cdot(z,\sigma,\tau)) {\ZZ}_S(T^{r}S^{-1}t_2^{-h}\cdot(z,\sigma,\tau)),\nonumber\\
\ZZ_{g_{r,h}}(z,\sigma,\tau)&=&\Gamma(z+\tau-h\sigma,\sigma,\tau)\Gamma(z-h\sigma,r\sigma-\tau,\sigma)\prod_{j=1}^{h}\theta(z-j\sigma,\tau),\nonumber\\
&=&\Gamma(z+\tau,\sigma,\tau)\Gamma(z-h\sigma,r\sigma-\tau,\sigma)\qquad ({\rm mod}\,\, \tM).
\eea
Here we have used ${\hat \ZZ}_{t_2^h}(z,\sigma,\tau)=\textstyle{\prod_{j=1}^{h}\theta(z-j\sigma,\tau)}$ which follows directly from equation \eqref{generators}. In order to get the expression quoted in \cite{Razamat:2013opa}, we  need to change $z~= ~{\tilde z}~+~h\ttau$, 
\be
\ZZ_{g_{r,h}}(z,\sigma,\tau)=\Gamma({\tilde z}+(r+h)\ttau,\tsigma+\ttau ,r\ttau)\Gamma({\tilde z}-h\tsigma,\tsigma+\ttau,r\tsigma)\qquad ({\rm mod}\,\, \tM).
\ee
As emphasised earlier, all the expressions obtained are modulo phases. This is a serious drawback if the phase factors depend on the holonomy of the gauge symmetry. Because, we are thinking of our modular constraints as constraints on the partition functions and not as constraints on the ``integrand" of the gauge holonomy integral, gauge holonomies never appears in our partition functions. 

\subsection{Holomorphic block factorization}
We have  discussed the presentation of the  supersymmetric partition function on background corresponding to  the gluing by element $g$ as the inner product,
\bea\label{factorization}
\ZZ_g^{\alpha\beta}(z,\sigma,\tau)&=&\delta_{\alpha\beta} \langle \alpha,(z,\sigma,\tau)|\alpha,g^{-1}\cdot(z,\sigma,\tau)\rangle = \delta_{\alpha\beta}\,\BB_{\tt L}^{\alpha}(z,\sigma,\tau) \, \BB_{\tt R}^{\alpha}(g^{-1}\cdot (z,\sigma,\tau)).\nonumber\\
\hat \ZZ_g^{\alpha}(z,\sigma,\tau)&=&\frac{\ZZ_g^{\alpha\alpha}(z,\sigma,\tau)}{\ZZ_1^{\alpha\alpha}(g^{-1}\cdot (z,\sigma,\tau))}=\frac{\BB_{\tt L}^{\alpha}(z,\sigma,\tau)}{\BB_{\tt L}^{\alpha}(g^{-1}\cdot (z,\sigma,\tau))}.
\eea
However, due to the curvature of the Berry connection this equation must be valid only locally i.e. only for some group elements $g$. If this equation were valid globally, then this would make $\hat \ZZ_g^{\alpha}(z,\sigma,\tau)$ a trivial element of $H^1({\cal G},\tN/\tM)$ and hence $e^{i\phi_{g_1,g_2}}$ appearing in equation \eqref{master} a trivial element of $H^2({\cal G},\tM)$. The Berry curvature is precisely the obstruction to that. 

In what follows, we will study the validity of this expression for all the generators of ${\cal G}$. The holomorphic blocks in four dimensions have been computed in \cite{Peelaers:2014ima, Nieri:2015yia, Longhi:2019hdh}. The right blocks $\BB_{\tt R}^\alpha$ are related to the left ones $\BB_{\tt L}^\alpha$ by orientation reversal. Concretely, ${\cal B}^{\mathtt R}_\alpha(z,\sigma,\tau)={\cal B}^{\mathtt L}_\alpha(z,-\sigma,\tau)$.
For free chiral it is known that,
\be
{\cal B}^{\mathtt L}(z,\sigma,\tau)= \Gamma(\frac{z}{\tau},\frac{\sigma}{\tau},-\frac{1}{\tau})\quad ({\rm mod}\,\,\tM).
\ee
From here we compute,
\bea
\ZZ_1(z,\sigma,\tau)&=&{\cal B}^{\mathtt L}(z,\sigma,\tau){\cal B}^{\mathtt R}(z,\sigma,\tau)=1/\theta(z,\tau),\nonumber\\
\ZZ_{t_2}(z,\sigma,\tau)&=&{\cal B}^{\mathtt L}(z,\sigma,\tau){\cal B}^{\mathtt R}(t_2^{-1}\cdot(z,\sigma,\tau))=1,\nonumber\\
\ZZ_{S_{23}}(z,\sigma,\tau) &=&{\cal B}^{\mathtt L}(z,\sigma,\tau){\cal B}^{\mathtt R}(S_{23}^{-1}\cdot(z,\sigma,\tau))=\Gamma(z,\sigma,\tau).
\eea
Happily, these expressions agree with the explicit localization computations given in equation \eqref{localization}. However, the situation is different for some $h\in {\cal H}$. Recall that we expect $\hat \ZZ_{h}=1$. If we use the formula \eqref{factorization}, it is straightforward to see that we get
\be
{\hat \ZZ}_{T_{13}}={\hat \ZZ}_{T_{23}}={\hat \ZZ}_{t_3}=1,  
\ee
but for the other generators of $\cal H$,
\be
{\hat \ZZ}_{S_{13}}=\frac{\Gamma(\frac{z}{\tau},\frac{\sigma}{\tau},-\frac{1}{\tau})}{\Gamma(z,\sigma,\tau)}\neq 1,\quad {\hat \ZZ}_{T_{21}}=\frac{\Gamma(\frac{z}{\tau},\frac{\sigma}{\tau},-\frac{1}{\tau})}{\Gamma(\frac{z}{\tau},\frac{\sigma-1}{\tau},-\frac{1}{\tau})}\neq 1,\quad {\hat \ZZ}_{t_{1}}=\frac{\Gamma(\frac{z}{\tau},\frac{\sigma}{\tau},-\frac{1}{\tau})}{\Gamma(\frac{z-1}{\tau},\frac{\sigma}{\tau},-\frac{1}{\tau})}\neq 1.
\ee
From this it is clear that the factorization formula \eqref{factorization} does not work for $S_{13},T_{21}, t_{1}$. As remarked earlier, this is to be expected. If the factorization formula worked for all elements of $\cal G$, $\hat \ZZ_{g}$ would  be trivial in cohomology which is not the case. We explicitly see here  how \eqref{factorization} is local trivialization of something that is non-trivial in cohomology.
We  find it striking that the issues having to do with Berry curvature end up making such a drastic impact on the partition functions.

Because the holomorphic block decomposition works for all elements of $\cal G$ except possibly for the ones involving $\{S_{13}, T_{21}, t_{1}\}$, in hindsight, all the relations in \eqref{relations} had to work. The relation that remains nontrivial is $Y^3=1$. 
This is because $\hat \ZZ_g$ is admits a local trivialization for all the generators of $\cal G$ involved in the relations except for $S_{13}$ (which is needed to construct $Y$). The fact that the relation $Y^3=1$ is also respected by the partition functions is the statement that $\hat \ZZ_g$ is a non-trivial element of the cohomology.

\section{Examples}\label{examples}
In this section, we will study the partition function for interacting supersymmetric theories.  First, we will look at the chiral multiplet with R-charge $R$ and then at supersymmetric $U(1)$ gauge theory.
We will be mainly interested in the constraint imposed on the partition functions by the group relation $Y^3=1$. For this purpose, we will need $\hat \ZZ_{Y}$. As discussed in equation \eqref{Y-S}, $\hat \ZZ_{Y}=\hat \ZZ_{S_{23}^{-1}}$. We compute this normalized partition function from the supersymmetric index using holomorphic block decomposition with the help of equation \eqref{factorization}.

\subsection{Chiral multiplet with R-charge}\label{chiral-R}
In what follows it is convenient to use homogeneous coordinates $(\omega_1\,\omega_2\,\omega_3)$ in addition to the affine coordinates $(1\,\sigma\,\tau)$ that we have been so far using. The simplest interacting ${\cal N}=1$ supersymmetric conformal field theory is that of chiral multiplets interacting through a superpotential. The effect of superpotential is to impart non-trivial R-charge to the chiral multiplet. 

Consider a chiral multiplet $\Phi$ interacting with the superpotential $\Phi^n$. This implies its R-charge is $R=2/n$. The theory also has a global symmetry ${\mathbb Z}_n$. Allowing for a fugacity for this global symmetry, we get the superconformal index of this theory to be 
\be
\ZZ_{S_{23}}(\sigma,\tau)=\Gamma(\frac{R}{2}+\frac{R}{2}(\sigma+\tau),\sigma,\tau)=\Gamma(\frac{R}{2}\frac{\omega_1+\omega_2+\omega_3}{\omega_1},\frac{\omega_2}{\omega_1},\frac{\omega_3}{\omega_1}).
\ee
This is same as the index of the chiral multiplet with R-charge $0$ with the background $U(1)$ holonomy $\xi\to R\os$, where we have defined $\os=\textstyle{\sum_1^3} \omega_i/2$. In more complicated theories, the chiral multiplet with non-trivial R-charge could also be charged under another background $U(1)$ global symmetry. In that case, the contribution of such a chiral multiplet to the superconformal index is 
\be
\ZZ_{S_{23}}(z,\sigma,\tau)=\Gamma(z+\frac{R}{2}+\frac{R}{2}(\sigma+\tau),\sigma,\tau)=\Gamma(\frac{\xi}{\omega_1}+\frac{R\os}{\omega_1},\frac{\omega_2}{\omega_1},\frac{\omega_3}{\omega_1}).
\ee
In other words, having nontrivial R-charge $R$ effectively shifts the global symmetry holonomy by $\xi\to \xi+R\os$. The shift of $\xi$ by $R\omega_1/2$ in addition to the standard shift by $(\omega_2+\omega_3)R/2$ can also be achieved by working with the so-called modified index \cite{Kim:2019yrz}. It entails replacing $(-1)^F$ in the supersymmetric index \eqref{4dindex} by $e^{i\pi R}$. Because the supercharge has R-charge $1$, the robustness properties of the Witten index are unaffected.

The holomorphic block is
\be
\BB_{\tt L}(z,\sigma, \tau)=\Gamma(\frac{z+\frac{R}{2}(\sigma+\tau+1)}{\tau},\frac\sigma\tau,-\frac1\tau)\quad({\rm mod}\,\, \tM)
\ee
From here it is not difficult to see that 
\bea
{\hat \ZZ}_{S_{23}^{-1}}(z,\sigma,\tau)&=&\frac{\BB_{\tt L}(z,\sigma,\tau)}{\BB_{\tt L}(S_{23}\cdot(z,\sigma ,\tau))}=\Gamma(\frac{z+\frac{R}{2}(\sigma+\tau+1)}{\tau},\frac\sigma\tau,-\frac1\tau) \Gamma(\frac{z+\frac{R}{2}(-\sigma+\tau+1)}{\sigma},\frac\tau\sigma,-\frac1\sigma)\nonumber\\
&=&{\hat \ZZ}_{Y}(z,\sigma,\tau).
\eea
Due to $R$ dependent terms the equation \eqref{modularY} 
implied by the group relation $Y^3=1$ is not obeyed by above $\hat \ZZ_Y$ for general values of $R$. But this is not surprizing. This is because, as pointed out in \cite{Closset:2013vra}, in order to preserve supersymmetry on $S^2\times {\mathbb T}^2$  R-charges must be quantized to be integers.  In fact, the theory needs to preserve supersymmetry on all lens spaces $L(r,s)$, for this the R-charges need to be quantized as even-integers.  If we stick to gluing by a subgroup of $SL(3,{\mathbb Z})$ that does not produce lens spaces, then it suffices for the R-charges to be integers. As the $SL(3,{\mathbb Z})$ element $Y$ can not to produce lens spaces under gluing, the relation $Y^3=1$ requires the R-charges to be only integers and not necessarily even-integers. This is what we will assume.

For integer R-charge,
\be
\Gamma(\frac{z+\frac{R}{2}(-\sigma+\tau+1)}{\sigma},\frac\tau\sigma,-\frac1\sigma)=\Gamma(\frac{z+\frac{R}{2}(\sigma+\tau+1)}{\sigma},\frac\tau\sigma,-\frac1\sigma).
\ee
This means 
\be
{\hat \ZZ}_{Y}(z,\sigma,\tau)=\Gamma(z+\frac{R}{2}(1+\sigma+\tau),\sigma, \tau)=\Gamma(\frac{\xi+R\os}{\omega_1},\frac{\omega_2}{\omega_1},\frac{\omega_3}{\omega_1})=\ZZ_{S_{23}}(z,\sigma,\tau) \quad({\rm mod}\,\,\tM).
\ee
Again, taking $\hat \ZZ_Y(z,\sigma,\tau)=\ZZ_{S_{23}}(z,\sigma,\tau)$ we see that it indeed obeys the equation \eqref{modularY}. This is because the shift in the holonomy $z$ is symmetric under $Y$. The function $P(z,\sigma,\tau)$ is given by
\bea\label{P-R}
P_{\chi_R}(z,\sigma,\tau)&=&Q(z+\frac{R}{2}(1+\sigma+\tau),\sigma, \tau)\\
&=&\frac{1}{\prod_i\omega_i}\Big(\xi^3+3\xi^2\os (R-1)+3 \xi\os^2 (R-1)^2+\os^3 (R-1)^3-\xi \oss-\os\oss(R-1)\Big).\nonumber
\eea
Recall the definitions $\os=\textstyle{\sum_i}\omega_i/2$ and $\oss=\textstyle{\sum_i} \omega_i^2/4$. Identifying $(R-1)^i={\rm Tr}\, R^i$  and  using global symmetry charge $F=1$, we see that $P(z,\sigma,\tau)$ is precisely the anomaly polynomial for the chiral multiplet with R-charge $R$.
\bea\label{phase-anomaly}
P_{\chi_R}(z,\sigma,\tau)&=&\frac{1}{\prod_i\omega_i}\Big(\xi^3\,k_{FFF}+3\xi^2\os \,k_{FFR}+3 \xi\os^2 \,k_{FRR}+\os^3 \,k_{RRR}\nonumber\\
&-&\xi \oss \,k_F-\os\oss\,k_R\Big).
\eea

\subsection{SQED in $4d$}\label{gauge-theory}
As in the previous section, we will compute $\hat \ZZ_Y$ by computing the holomorphic blocks which in turn are computed using the index $\ZZ_{S_{23}}$ and then factorizing it. Factorization of  the supersymmetric index has been studied in  \cite{Peelaers:2014ima, yoshida2014factorization, Nieri:2015yia}. One important difference from the case of only chiral multiplets is that the supersymmetric gauge theory has multiple Higgs branch vacua. Partition function in each vacuum can be dealt with separately because the equation \eqref{master} is diagonal. As before, the group relation $Y^3=1$ yields non-trivial constraints that must be satisfied by $\hat \ZZ^\alpha_Y$'s.

The index of a $U(1)$ gauge theory with $N$ flavors is given by the integral
\bea
\ZZ_{S_{23}}({\vec \alpha},{\vec \beta},\sigma,\tau)&=&(\sigma,\sigma)(\tau,\tau)\oint d\xi \prod_{j=1}^{N}\Gamma (\frac{R(\sigma+\tau)}{2}+\xi+\alpha_j,\sigma,\tau)\nonumber\\
&\times & \Gamma(\frac{R(\sigma+\tau)}{2}+\beta_j-\xi,\sigma,\tau).
\eea
Here $R$ is the $U(1)_R$ charge of the chiral multiplet. For anomaly cancellation we need $R=1$. The variables $\alpha,\beta$ are $SU(N)$ holonomies and hence obey $\Sigma_i\alpha_i=\Sigma_i\beta_i=0$. The function $(z,\tau)$ is called the Pochhammer symbol defined as $(z,\tau)\equiv \textstyle{\prod_{n=1}^{\infty}}(1-x q^i)$ where $x=\exp(2\pi i z), q=\exp(2\pi i \tau)$. The flavor symmetry $F=SU(N)$ that acts only on chirals with gauge charge $+1$ (or chirals with gauge charge $-1$) has an non-zero $FFG$ anomaly with the gauge symmetry $G=U(1)$. We will only turn on background holonomies that do have an anomaly with the gauge symmetry. The global symmetry that does not have anomaly with the gauge symmetry is the diagonal combinations of the two $SU(N)$ symmetries. In order to turn on holonomies only for this symmetry we must set  $\beta_i=\alpha_i$. The abelian symmetry $F'$ that acts on all the chiral multiplets also has an anomaly with the gauge symmetry. This anomaly is a more standard $F'GG$ type ABJ anomaly. We do not turn on background holonomy for this abelian global symmetry. 

This integral is evaluated by summing over all the poles inside the unit circle, it can be written in a sum over Higgs branch vacua where summand is a factorized product. Picking the poles coming form negatively charged chiral multiplets,
\bea\label{gauge-index}
\ZZ_{S_{23}}&=&\sum_{j=1}^{N}\prod_{i}\Gamma(\sigma+\tau+\beta_{i}+\beta_{j},\sigma, \tau)\prod_{i\neq j}\Gamma(\beta_i-\beta_j,\sigma,\tau)Z_V^{(j)}({\vec \beta};\sigma,\tau)
Z_V^{(j)}({\vec \beta};\tau,\sigma),\nonumber\\
\ZZ_{S_{23}}&=&\frac{1}{\Gamma(0,\sigma,\tau)}\sum_{j=1}^{N}\prod_{i}\frac{\Gamma(\beta_i-\beta_j,\sigma,\tau)}{\Gamma(-\beta_{i}-\beta_{j},\sigma, \tau)}Z_V^{(j)}({\vec \beta};\sigma,\tau)
Z_V^{(j)}({\vec \beta};\tau,\sigma),\nonumber\\
\ZZ_{S_{23}}^{jj}&=&\frac{1}{\Gamma(0,\sigma,\tau)}\prod_{i}\frac{\Gamma(\beta_i-\beta_j,\sigma,\tau)}{\Gamma(-\beta_{i}-\beta_{j},\sigma, \tau)}Z_V^{(j)}({\vec \beta};\sigma,\tau)
Z_V^{(j)}({\vec \beta};\tau,\sigma),\nonumber\\
&\equiv&Z_P^{(j)}({\vec \beta};\sigma,\tau)Z_V^{(j)}({\vec \beta};\sigma,\tau)
Z_V^{(j)}({\vec \beta};\tau,\sigma)
\eea
In the second line we have used $\Gamma(z)=1/\Gamma(\sigma+\tau-z)$, we have formally included $i=j$ term in the second set of product which is infinity and have divided by $\Gamma(0)$ to get rid of it. We have done it because it is then uniform for all $i$s. We have defined the ``vortex partition function",
\be
Z_V^{(j)}({\vec \beta};\sigma,\tau):= \sum_{s=0}^{\infty}\prod_{i}\frac{\Theta(\sigma+\beta_{i}+\beta_{j};\sigma;\tau)_s}{\Theta(\sigma+\beta_j-\beta_i;\sigma;\tau)_s},
\ee
where $\Theta(z;\sigma;\tau)_n$ is the theta factorial defined in equation \eqref{theta-fac}. A quick glance at equation \eqref{ehs} indicates that the vortex partition function is in fact an elliptic hypergeometric series.
\be
Z_V^{(j)}({\vec \beta};\sigma,\tau)=\,_N E_{N-1}(\sigma+\vec\beta+\beta_{j}, \sigma+\beta_{j}-\vec\beta; \sigma;\tau;1).
\ee
Interestingly, this elliptic hypergeometric series is invariant under transformation $S_{13}$. This can be checked explicitly by making a modular transformation for each of the theta function involved in the elliptic hypergeometric series. The arguments of the theta functions involved in the product are such that all the phases coming from the modular transformation cancel rendering each term of the sum separately modular invariant.
It is worth noting that the vortex partition function $Z_V$ can be obtained from the perturbative part $Z_P$ with the action of a simple difference operator.
\be\label{vortex-diff}
Z_V(\vec\beta,\sigma,\tau)=\frac{1}{Z_P(\vec\beta,\sigma,\tau)} \Big(\frac{1}{1-t^{\beta_j}_2}\,Z_P(\vec\beta,\sigma,\tau)\Big).
\ee

The $j$-th holomorphic block modulo $\tM$ is
\be
{\cal B}^{(j)}({\vec \beta};\sigma,\tau)= Z_P^{(j)}(\frac{{\vec \beta}}{\tau};\frac{\sigma}{\tau},-\frac{1}{\tau}) \times  Z_V^{(j)}({\vec \beta};\sigma,\tau)\qquad ({\rm mod}\,\,\tM).
\ee
Here we have used the fact that $Z_P^{(j)}$ is a product of elliptic gamma functions and we have factored it by factoring each elliptic gamma function as,
\be
\Gamma(z,\tau,\sigma)=\Gamma(\frac{z}{\tau},\frac{\sigma}{\tau},-\frac{1}{\tau})\Gamma(\frac{z}{\sigma},\frac{\tau}{\sigma},-\frac{1}{\sigma})\qquad ({\rm mod}\,\,\tM).
\ee
It is convenient to think of the holomorphic block as the perturbative part $Z_P^{(j)}$ consisting of product of elliptic gamma function and the vortex part $Z_V^{(j)}$ consisting of the elliptic hypergeometric series.

Finally, we write down the normalized partition function $\hat \ZZ_Y$ using equation \eqref{factor},
\bea\label{zzy}
{\hat \ZZ}_{Y}^{(j)}({\vec \beta};\sigma,\tau)&=&{\hat \ZZ}_{S_{23}^{-1}}^{(j)}({\vec \beta};\sigma,\tau)=\frac{{\cal B}^{(j)}({\vec \beta};\sigma,\tau)}{{\cal B}^{(j)}({\vec \beta};\tau,-\sigma)}=Z_P^{(j)}({\vec \beta};\sigma,\tau)\frac{Z_V^{(j)}({\vec \beta};\sigma,\tau)}{Z_V^{(j)}(S_{23}\cdot({\vec \beta};\sigma,\tau))}\qquad ({\rm mod}\,\,\tM),\nonumber\\
&=&Z_P^{(j)}({\vec \beta};\sigma,\tau)\frac{Z_V^{(j)}({\vec \beta};\sigma,\tau)}{Z_V^{(j)}(Y^{-1}\cdot({\vec \beta};\sigma,\tau))}\qquad ({\rm mod}\,\,\tM).
\eea
Here also we have used the $S_{13}$ invariance of the vortex partition function $Z_V^{(j)}$. From the final form it is clear that the vortex contribution to ${\hat \ZZ}_Y$ is cohomologically trivial. 

Defining $\hat \ZZ_Y$ with equation \eqref{zzy} and dropping $(\rm mod)\,\,\tM$, we evaluate the equation \eqref{modularY}.
\bea
\hat \ZZ_Y^{(j)}({\vec \beta};\sigma,\tau)\hat \ZZ_Y^{(j)}(Y^{-1}\cdot ({\vec \beta};\sigma,\tau))\hat \ZZ_Y^{(j)}(Y^{-2}\cdot ({\vec \beta};\sigma,\tau))&=&e^{-i\frac{\pi }{3}P_{\rm SQED}({\vec \beta};\sigma,\tau)},\nonumber\\
\Rightarrow \quad Z_P^{(j)}({\vec \beta};\sigma,\tau)Z_P^{(j)}(Y^{-1}\cdot ({\vec \beta};\sigma,\tau))Z_P^{(j)}(Y^{-2}\cdot ({\vec \beta};\sigma,\tau))&=&e^{-i\frac{\pi }{3}P_{\rm SQED}({\vec \beta};\sigma,\tau)}.
\eea
Note that the vortex partition function part $Z_V$ is totally cancelled from this equation. This is because its contribution is cohomologically trivial. The constraint is only on the perturbative part $Z_P$ which is product of elliptic gamma functions. 
The phase $P_{\rm SQED}(\vec\beta,\sigma,\tau)$ is computed by summing all the phases induced by individual elliptic gamma functions.
\bea
{P}_{\rm SQED}(\vec\beta,\sigma,\tau)&=&-{Q}(0,\sigma,\tau)+\sum_i\Big({Q}(\beta_i-\beta_j,\sigma,\tau)-{Q}(-\beta_i-\beta_j,\sigma,\tau)\Big),\nonumber\\
&=& 2\sum_i \beta_i^3+\os^3-\os\oss.
\eea
This is indeed the 't-Hooft anomaly polynomial \eqref{anomalyY}. Anomalies of the theory can be read off easily from the field content: fundamental and anti-fundamental chiral multiplets with R-charge $1$ and a vector multiplet (it has a gaugino with R-charge $2$). This yields
\be
k_{FFF}=2,\quad k_{FFR}=0,\quad k_{FFR}=0,\quad k_{RRR}=1,\quad k_F=0,\quad k_R=1.
\ee
Here $F$ stands for the $SU(N)$ global symmetry.

To end this section, we would like to point out some properties of the partition functions that we expect to be true for any ${\cal N}=1$ supersymmetric gauge theory. The supersymmetric index for any gauge theory is given as an integral over gauge holonomies of a product of elliptic gamma function. After evaluating the integral by summing the residues of all the poles, the partition function schematically takes the form,
\be
\ZZ_{S_{23}}({z_i};\sigma,\tau)=\sum_\alpha Z_P^{\alpha}({z_i};\sigma,\tau)Z_V^{\alpha}({z_i};\sigma,\tau)
Z_V^{\alpha}({z_i};\tau,\sigma)
\ee
Here $z_i$ stand for all the global symmetry holonomies,  $Z_P$ is the perturbative part that is product of elliptic gamma functions and $Z_V$ is an elliptic hypergeometric series that is exactly modular invariant i.e. invariant under $S_{13}$. Moreover, thanks to the integrality of R-charges, $\hat \ZZ_Y$ also has the same form as in equation \eqref{zzy} i.e.
\be
\hat \ZZ_Y^{\alpha}({z_i};\sigma,\tau)=Z_P^{\alpha}({z_i};\sigma,\tau)\frac{Z_V^{\alpha}({z_i};\sigma,\tau)}{Z_V^{\alpha}(Y^{-1}\cdot ({z_i};\sigma,\tau))}\qquad({\rm mod} \,\,\tM).
\ee
Equation \eqref{modularY} is evaluated by defining $\hat \ZZ_Y^\alpha$ to be the right hand side of the above equation with $({\rm mod} \,\,\tM)$ dropped. We will use these properties of the gauge theory partition functions to formulate a bootstrap program in section \ref{bootstrap}.

For ${\cal N}=2$ supersymmetric gauge theories, holomorphic blocks on one side trivialize in the Schur limit of the superconformal index. This observation may be important in establishing a connection with the $SL(2,{\mathbb Z})$ modular property of the Schur index discovered in \cite{Razamat:2012uv}.

\section{Group cohomology}\label{global-grav}

In this section, we will give a quick review of group cohomology. We will follow the short but excellent introduction to the topic given in \cite{Dijkgraaf:1989pz}. To understand group cohomology, it is useful to introduce the notion of classifying space. The classifying space $BG$ is the base space of the principal $G$ bundle $EG$, the so-called universal bundle. 
Any $G$ bundle on manifold $W$ is classified by the so-called ``classifying map" $\gamma: W\to BG$. The topology of the bundle $E$ is completely determined by the homotopy of the classifying map. In general, the classifying space of a compact group is an infinite-dimensional space. The group cohomology of $G$ is nothing but the cohomology of the classifying space $BG$. Although it is the cohomology of the classifying space, the group cohomology valued in the $G$ module $M$ is usually denoted as $H^*(G, M)$. We will stick to this convention as the cohomology of the group itself, as a topological space, will not play an important role for us\footnote{The groups that interest us are discrete.}. Given an element of the group cohomology, its pullback under the classifying map yields an element of $H^*(W,M)$ which only depends on the $G$ bundle $E$.

\begin{figure}[h]
    \centering
    {\includegraphics[width=0.6\textwidth]{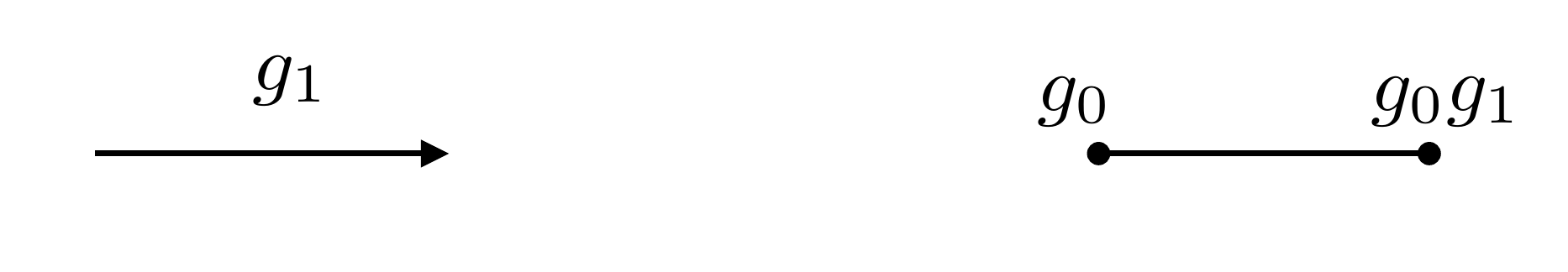} } \\
     {\includegraphics[width=0.66\textwidth]{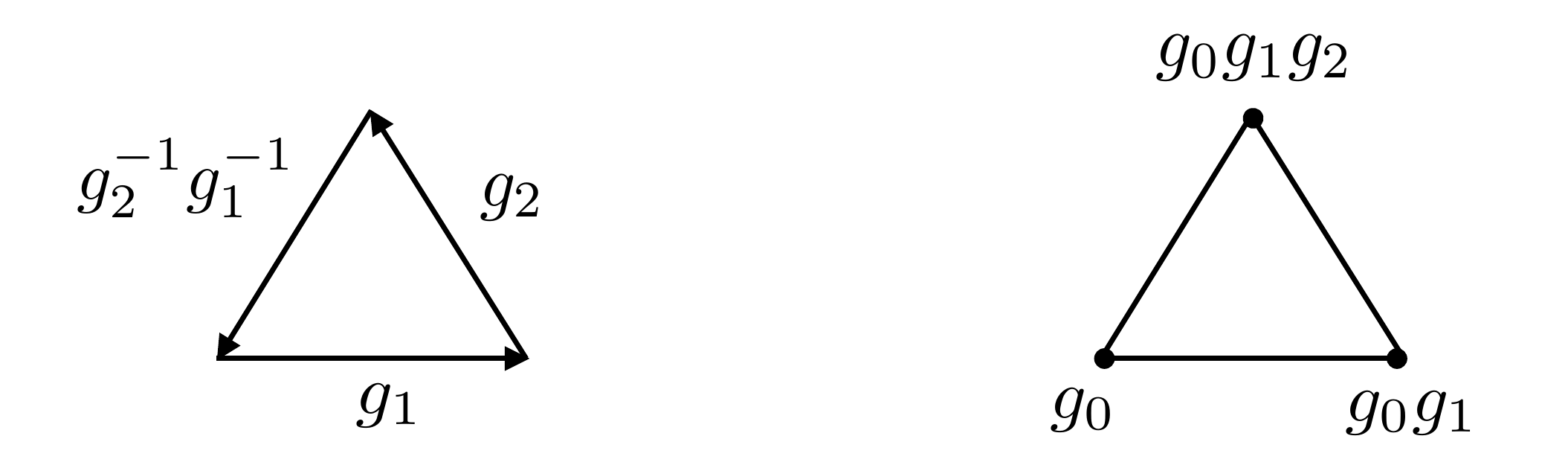} } 
    \caption{We have shown a 1-simplex and a 2-simplex in $BG$. In the first column, the simplices are denoted as $\alpha(g_1)$ and $\alpha(g_1,g_2)$ respectively while in the second column they are denoted as $\nu(g_0,g_0g_1)$ and $\nu(g_0,g_0g_1,g_0g_1g_2)$ respectively.}\label{simplices}
\end{figure}

For discrete groups, the group cohomology has an algebraic description. A $1$-simplex in $BG$ is labeled by a single group elements, a $2$-simplex in $BG$ is labeled by two group elements and so on. Naturally, the group cochain $C^k(G,M)$  is a map $\alpha:G^k\to M$. It turns out that the coboundary operator given by
\bea
(\delta\alpha)(g_1,\ldots, g_{k+1})&=&\alpha(g_1,\ldots, g_k) \,[(g_1\cdot \alpha(g_2,\ldots,g_{k+1}))\nonumber\\
&\times &\prod_{i=1}^k\alpha(g_1,\ldots, g_{i} g_{i+1},\ldots, g_{k+1})^{(-1)^i}]^{(-1)^{k+1}}.
\eea
The formula becomes more transparent if we use a reference ``point" $g_0$ as the $0$-simplex. Then a $1$-simplex is given by the two vertices $g_0$ and $g_1$, a $2$-simplex is given by the three vertices $\{g_0,g_1,g_2\}$ and so on. With this notation, the group cochain $C^k(G,M)$  is a ``homogeneous" map $\nu:G^{k+1}\to M$ i.e. $g\cdot \nu(g_0,\ldots,g_k)=\nu(gg_0,\ldots,gg_k)$. The function $\alpha(g_1,\ldots, g_{k})=\nu(1,\tilde g_1,\ldots, \tilde g_{k})$, $\tilde g_i=g_1\ldots g_i$.
The coboundary operator is
\be
(\delta\nu)(g_0,\ldots,g_{k+1})=\prod_{i=0}^{k+1}\nu(g_0,\ldots, {\hat g}_i,\ldots, g_{k+1})^{(-1)^{i+k+1}}.
\ee 
As usual, argument with the hat means that it is omitted. The group cohomology is $H^k(G,M)=Z^k(G,M)/B^{k}(G,M)$ where $Z^k(G,M)$ are the cocycles i.e. $\delta\alpha_k=0$ and $B^k(G,M)$ are the coboundaries i.e. $\alpha_k=\delta(\ldots)$. In figure \ref{simplices} we have shown a 1-simplex and a 2-simplex in $BG$.

Group cohomology has been applied to classify Chern-Simons theories with a general compact gauge group $G$ in \cite{Dijkgraaf:1989pz}. A group that is not connected or simply connected can admit non-trivial principal bundle and for such bundles, the conventional definition of the theory in terms of Chern-Simons functional is not applicable as the gauge connection can not be thought of as a Lie-algebra valued one form. In such cases, group cohomology $H^3(G,U(1))$ serves to classify Chern-Simons theories. Thanks to the short exact sequence,
\be
0\to {\mathbb Z}\to{\mathbb R}\to  U(1)\to 0,
\ee
We have group cohomology isomorphism $H^3(G,U(1))\simeq H^4(G,{\mathbb Z})$. 

The application of group cohomology is most striking for when $G$ is a discrete group. Chern-Simons theory on a three-manifold $W$ is defined by assigning an action $e^{iS[A]}$ to a $G$-bundle. This is because once the action for a single connection in a given $G$-bundle is given, it can be computed using a standard method \cite{Dijkgraaf:1989pz} for other connections in that bundle. Principal $G$-bundles for a discrete group is specified by the unique flat connection that it admits. It is given by the map $\lambda:\pi_1(W)\to G$. This also specifies the classifying map. The $U(1)$ valued action functional with required physical properties is then specified by  $H^3(G,U(1))$.

Recently group cohomology has also been applied in classifying symmetry protected topological phases \cite{Chen:2011pg}. This application is closer in spirit to the application of the group cohomology that we have in this paper. In \cite{Chen:2011pg}, the authors find that the topological phase protected by symmetry $G$ in $d+1$ dimensions is classified by\footnote{We assume that $G$ does not contain time-reversal symmetry. In case it does, the G-module $U(1)$ with trivial $G$ action changes to $U(1)_T$. Here $U(1)_T$ is the G-module in which time-reversal symmetry $T$ acts by inversion.} $H^{d+1}(G,U(1))$. The case with $d=1$  is closest to our setup. A topological theory in $1+1$ dimension has dynamical degrees of freedom living only on the edges.  As one moves in the parameter space, the state at living at the edge gets a Berry phase even as we come back to the same point in the parameter space after doing a monodromy associated to the $G$ action. The Berry connection, in general, could have a curvature which is measured by $H^2(G,{\mathbb Z})\simeq H^1(G,U(1))$ \cite{Chen:2011pg}. It classifies the G-protected topological phases. The group cohomology class $H^2(G,{\mathbb Z})$ also classifies topological $\theta$-terms in two dimensions. The cocycles $\phi_{g_1,g_2}$ are computed by evaluating these $\theta$-terms on G-bundle on a disc corresponding to the classifying map given by the image of the 2-simplex in $BG$ as specified in figure \ref{simplices}.

\begin{figure}[h]
    \centering
    {\includegraphics[width=0.5\textwidth]{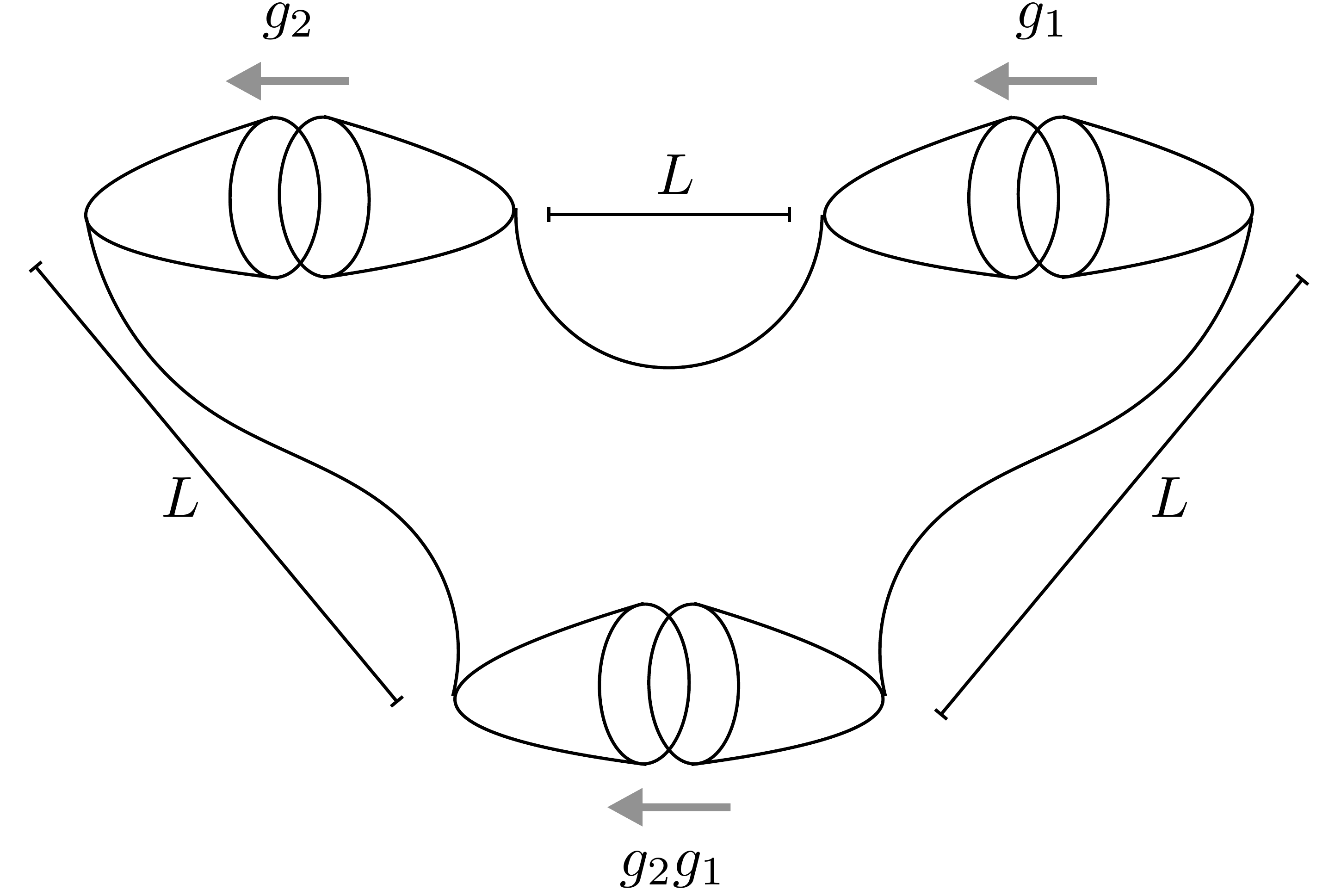} } 
    \caption{A cobordism between the disjoint sum of ${\cal M}_{g_1}$, ${\cal M}_{g_2}$ and ${\cal M}_{g_2g_1}$ is shown. It can be thought of as a ${\mathbb T}^3$ bundle on a 2-disc. In the limit the indicated segments of length $L$ become vanishingly small, the base space resembles the two-simplex of figure \ref{simplices}. The segments $L$ become the vertices.}\label{cobordism}
\end{figure}

In the case of supersymmetric partition functions in four dimensions that we have been working with, it may seem that the relevant cohomology is $H^4({\cal G},\tN/\tM)$  as we are working with a four-dimensional theory (the way it was $H^3(G,U(1))$ for Chern-Simons theory) but our problem is really one-dimensional. We are thinking of the four-dimensional manifold as a fibration of ${\mathbb T}^3$ on an interval. Hence the relevant $\cal G$ bundle is in fact on this interval. That's why the partition functions are classified by $H^1({\cal G},\tN/\tM)\simeq H^2({\cal G},\tM)$. We expect that $H^2({\cal G}, \tM)$ classifies Chern-Simons terms in five dimensions (five manifold is thought of as a ${\mathbb T}^3$ fibration). This is precisely the Chern-Simons anomaly polynomial. Then the 2-cocycles $\phi_{g_1,g_2}$  appearing in equation \eqref{master} are computed by evaluating the Chern-Simons terms on the associated ${\mathbb T}^3$ bundle on a disc. 
As before this bundle corresponds to the classifying map given by the image of the 2-simplex in $BG$ as specified in figure \ref{simplices}. Geometrically, this is five manifold is a cobordism that takes the disjoint sum of four manifolds ${\cal M}_{g_1}$ and ${\cal M}_{g_2}$ to ${\cal M}_{g_2g_1}$ as displayed in figure \ref{cobordism}.
 Here, by ${\cal M}_g$, we mean the four manifold obtained by gluing the solid three-tori by $g\in {\cal G}$. It would be extremely interesting to compute these 2-cocycles explicitly. We will leave this problem for the future. We should point out that for  the case of chiral multiplet of R-charge $0$, the 2-cocycles $\phi_{g_1,g_2}(z,\sigma,\tau)$ are explicitly given in \cite{Felder_2000}. It would be nice to match the result obtained from Chern-Simons with this.

\begin{figure}[h]
    \centering
    {\includegraphics[width=0.5\textwidth]{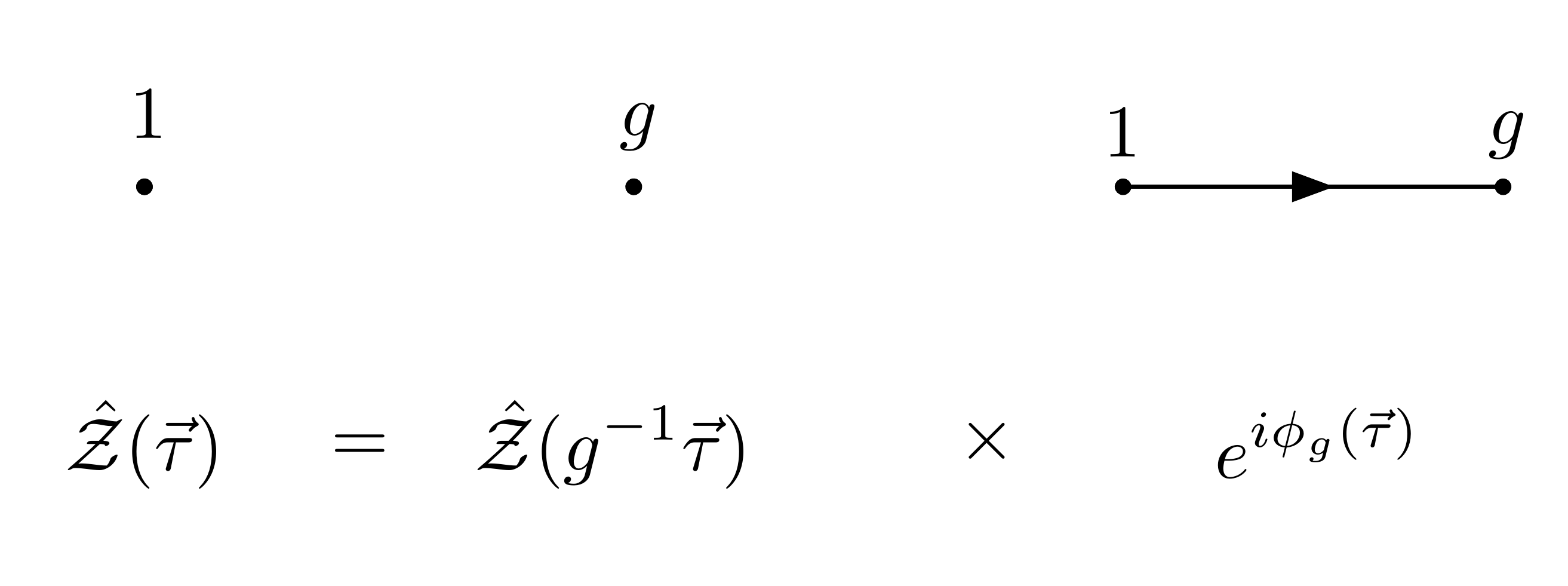} } 
    \caption{In two dimensions, the manifolds of interest are ${\mathbb T}^2$ fibration over a point i.e. just ${\mathbb T}^2$. In this figure, we have made the connection of the modular property of the ${\mathbb T}^2$ partition function with the group cohomology using a graphical notation. The partition function comes from the 0-simplex while the phases come from the 1-simplex as indicated. The group element $g$ stand for the group of large symmetries which contain the large diffeomorphism group $SL(2,{\mathbb Z})$ of ${\mathbb T}^2$ along with the large gauge transformations.}\label{h0general}
\end{figure}

\begin{figure}[h]
    \centering
    {\includegraphics[width=0.65\textwidth]{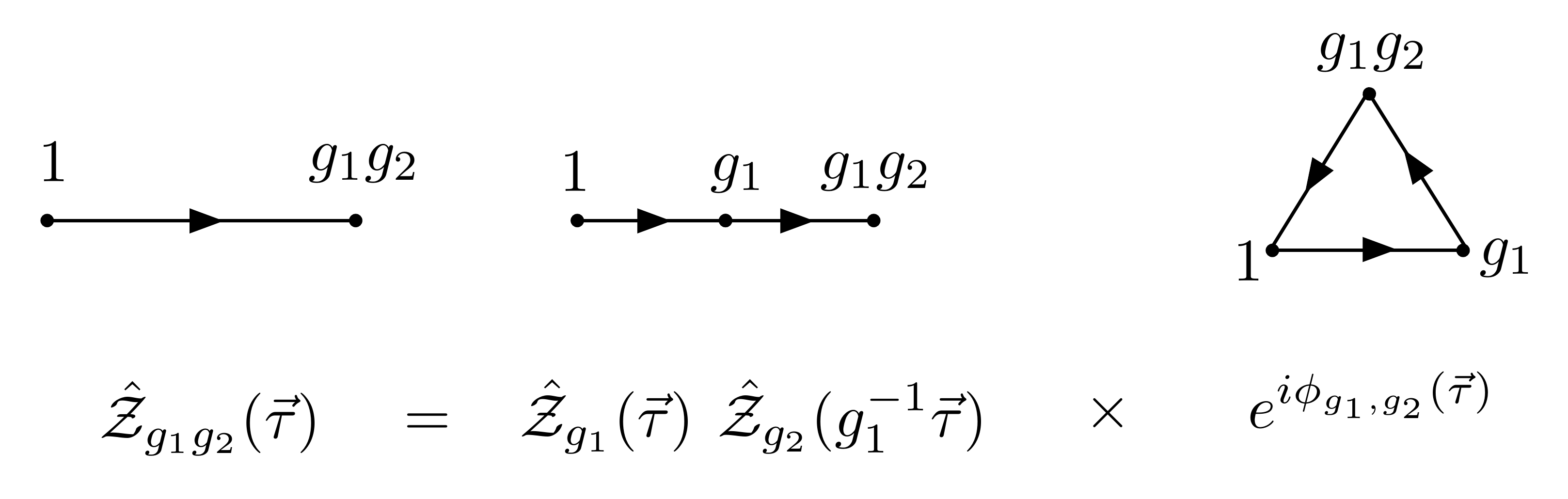} } 
    \caption{In four dimensions, the manifolds of interest are ${\mathbb T}^3$ fibration over an interval. In this figure, we have made the connection of the modular property of the four dimensional partition function \eqref{master} with the group cohomology using a graphical notation. The partition function comes from the 1-simplex while the phases come from the 2-simplex as indicated. The group elements $g_i$ stand for the group of large symmetries which contain the large diffeomorphism group $SL(3,{\mathbb Z})$ of ${\mathbb T}^3$ along with the large gauge transformations.}\label{h1general}
\end{figure}

\begin{figure}[h]
    \centering
    {\includegraphics[width=0.65\textwidth]{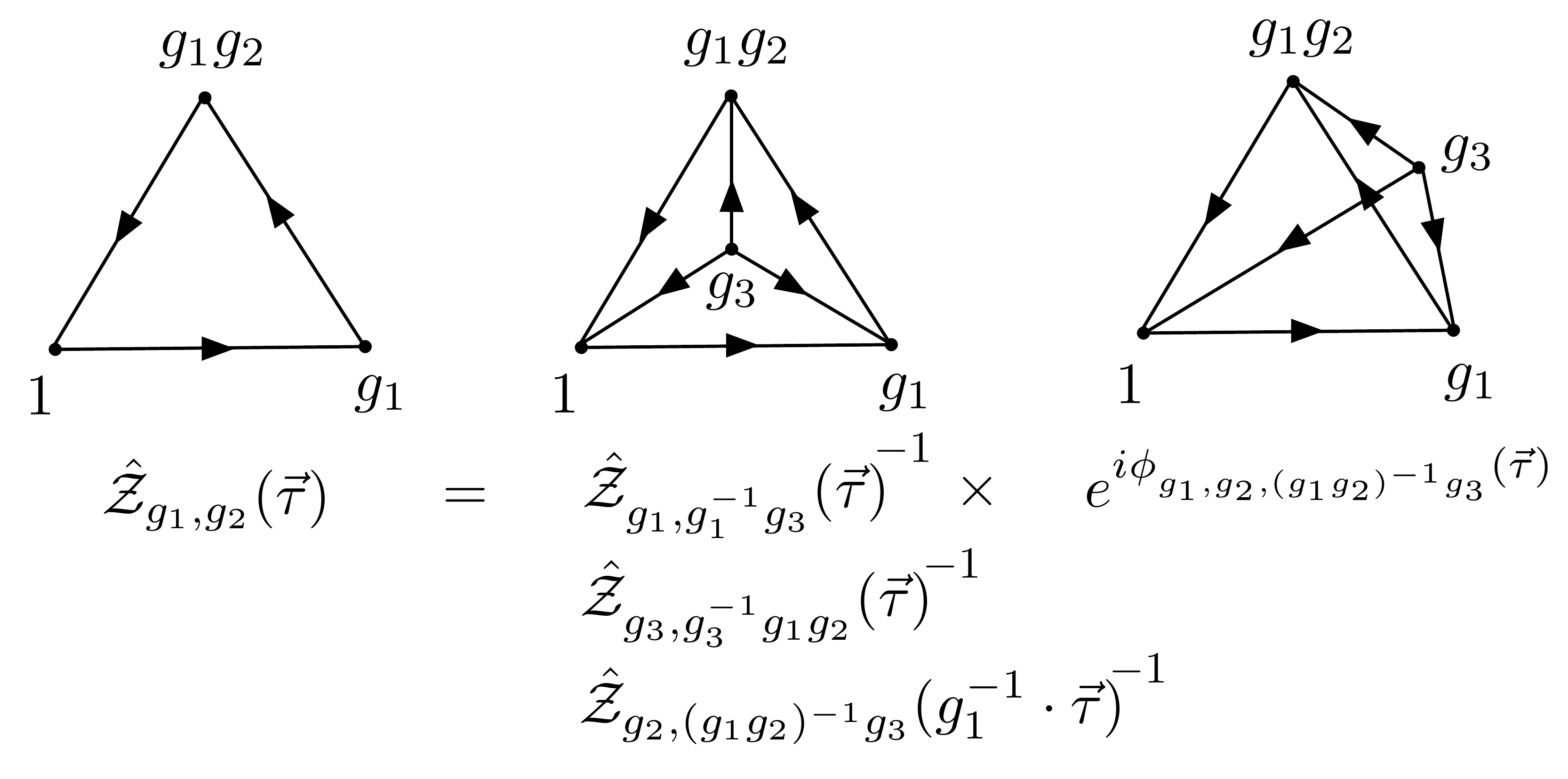} } 
    \caption{In six dimensions, the manifolds of interest are ${\mathbb T}^4$ fibration over a disc. In this figure, we have made the connection of the modular property of the four dimensional partition function \eqref{master} with the group cohomology using a graphical notation. The partition function comes from the 2-simplex while the phases come from the 3-simplex as indicated. The group elements $g_i$ stand for the group of large symmetries which contain the large diffeomorphism group $SL(4,{\mathbb Z})$ of ${\mathbb T}^4$ along with the large gauge transformations. The powers $\pm1$ of cocycles $\hat \ZZ$ come from if the even/odd number of arrows need to be flipped to get to the defining arrow configuration.}\label{h2general}
\end{figure}

Figure \ref{h0general} and \ref{h1general} give a graphical way of understanding  the 0-cocycle and 1-cocycle condition that is satisfied by the two-dimensional partition function \eqref{genus} and the four-dimensional partition functions \eqref{master} respectively. Figure \ref{h0general} describes a fibration of ${\mathbb T}^2$ on a point by specifying the classifying map i.e. the image of the base into $B{\cal G}_{2d}$ while figure \ref{h1general} describes a fibration of ${\mathbb T}^3$ on an interval by specifying the image of the base into $B{\cal G}$.
We have used $\nu$ notation (rather than $\alpha$ notation) to denote these cocycles graphically. 
These figures suggest a generalization of the modular properties of the two dimensional and four dimensional supersymmetric partition functions to supersymmetric partition functions of six dimensional theories on ${\mathbb T}^4$ fibrations over a disc. This tempting conjecture is summarized in figure \ref{h2general}. We will discuss more about this generalization and its consequences in section \ref{six-d}.

An interesting thing to note is that in the applications of group cohomology in \cite{Dijkgraaf:1989pz} and \cite{Chen:2011pg}, the coefficient system is a trivial G-module, however, in the problem at hand, the cohomology is valued in non-trivial G-modules $\tN$ and $\tM$. This makes the problem richer by making the explicit computation of cohomology rather non-trivial.

\section{Applications and discussion}\label{applications}
In this section we will discuss two applications of the constraints \eqref{master} satisfied by the supersymmetric partition functions and also a generalization of our results to supersymmetric theories in six dimensions. We will end with some outlook.

\subsection{A bootstrap program}\label{bootstrap}
Apart from normalized supersymmetric  partition functions $\hat \ZZ^\alpha_g$, equation \eqref{master} also contains the 2-cocycles $\phi_{g_1,g_2}$. As we have discussed in section \ref{global-grav}, one should be able to determine these functions explicitly by evaluating anomaly Chern-Simons form on the suitable cobordism such as the one given in figure \ref{cobordism}. Equation \eqref{master} when evaluated on group relations, yields constraints on $\hat \ZZ^\alpha_g$ of theories with a given 't-Hooft anomaly polynomial.
The analogous problem in two dimensions is the problem of classifying elliptic genera of a $(0,2)$ supersymmetric theory with given 't-Hooft anomaly polynomial using equation \eqref{genus}. This anomaly polynomial fixes the 1-cocycle appearing equation \eqref{genus}\footnote{By evaluating the anomaly Chern-Simons 3-form on a mapping cylinder.}.
In the case of $(2,2)$ supersymmetric theories, with only the holonomy for R-symmetry turned on, this problem reduces to the classic problem of classifying Jacobi forms of a given index (and zero weight). This index  is exactly the anomaly in R-symmetry.

We believe that non-trivial constraints on four-dimensional $\hat \ZZ^\alpha_g$ are obtained only due to the group relation $Y^3=1$. Restricting ourselves to only this constraint, the bootstrap program is given by the equation \eqref{modularY}. For SQED, we have shown that in equation \eqref{modularY}, $\hat \ZZ_Y^\alpha$ can be replaced by the perturbative part of the supersymmetric index $Z_P^\alpha$.  We have asserted that this is true for general gauge theories. This yields,
\begin{shaded}
\be\label{modularYY}
 {Z}_P^\alpha(z_i,\sigma,\tau) {Z}_P^\alpha(\frac{z_i}{\sigma},\frac{1}{\sigma},\frac{\tau}{\sigma}) {Z}_P^\alpha(\frac{z_i}{\tau},\frac{\sigma}{\tau},\frac{1}{\tau}) = e^{-i\frac{\pi}{3}P(z_i,\sigma,\tau)},
\ee
\end{shaded}
\noindent where the phase appearing on the right hand side $P(z_i,\sigma,\tau)$ is the anomaly polynomial in equation \eqref{anomalyY}. 
We would like to think of this equation in the same way as we think of the equation \eqref{genus} for the elliptic genus. For two-dimensional theories, equation \eqref{genus} classifies the elliptic genera of the theories with given the anomaly polynomial. The above equation lets us do the same for supersymmetric index in four dimension. Because the supersymmetric index is RG flow invariant, we think of this as a modular bootstrap equation for four-dimensional superconformal theories. 

Among all the solutions of equation \eqref{modularYY}, gauge theory solutions can be obtained by  assuming that $Z_P$ is a product of elliptic gamma functions.  With this assumption, we expect that the solution space of equation \eqref{modularYY} is tractable.
As remarked earlier, equation \eqref{modularYY} does not let us fix the vortex part $Z_V$. As $Z_V$ contribution is cohomologically trivial, cohomological considerations will not help us fix it. Once we know $Z_P$ that solves the above modular bootstrap equation \eqref{modularYY}, we expect difference operators such as the one in equation \eqref{vortex-diff} should let us fix $Z_V$ too. As is clear from the discussion, this tantalizing bootstrap program needs much more exploration. We plan to do so in the future.

 \subsection{A Cardy formula}
 We can study the ``high temperature" behavior of the supersymmetric index of gauge theories using equation \eqref{modularYY}. This is done by taking the limit $\sigma,\tau\to 0^{+i}$ i.e. $\sigma$ and $\tau$ approach $0$ along  positive imaginary axis. In this limit, the first term of the product is the high temperature limit of the perturbative partition function. The limit for the second and third term of the product is slightly subtle. Let's look at it bit closely for the case of chiral multiplet with R-charge $R$. In this case,
 \be
 \ZZ_P(z,\sigma,\tau)=\Gamma(z+\frac{R}{2}(1+\sigma+\tau),\sigma,\tau).
 \ee
 Note that this function is periodic under $z\to z+1$. We will compute its high temperature behavior in a suitable $z$ window of width $1$. The behavior at other points in $z$ is fixed due to the above periodicity.
 
 In the limit $\sigma,\tau\to 0^{+i}$, the second term of the product becomes 
 \bea
 \Gamma(\frac{z+\frac{R}{2}(\sigma+\tau+1)}{\sigma},\frac1\sigma,\frac\tau\sigma)&=&\frac{1}{ \Gamma(\frac{z-1+\frac{R}{2}(\sigma+\tau+1)}{\sigma},-\frac1\sigma,\frac\tau\sigma)}\\
 &=&\exp\Big(-\sum_{n=1}^\infty\frac{e^{n2\pi i \frac{z-1+\frac{R}{2}(\sigma+\tau+1)}{\sigma}}-e^{n2\pi i \frac{\tau-z-\frac{R}{2}(\sigma+\tau+1)}{\sigma}}}{(1-e^{-n2\pi i/\sigma })(1-e^{n2\pi i \tau/\sigma})}\Big)\nonumber\\
 &\xrightarrow{\sigma,\tau\to 0^{+i}}&  \exp\Big(-\sum_{n=1}^\infty\frac{e^{n2\pi i \frac{z-1+\frac{R}{2}}{\sigma}}-e^{n2\pi i \frac{-z-\frac{R}{2}}{\sigma}}}{(1-e^{n2\pi i \tau/\sigma})}\Big)
\eea
Similarly, the third term simplifies to,
\be
\Gamma(\frac{z+\frac{R}{2}(\sigma+\tau+1)}{\tau},\frac\sigma\tau,\frac1\tau)\xrightarrow{\sigma,\tau\to 0^{+i}}  \exp\Big(-\sum_{n=1}^\infty\frac{e^{n2\pi i \frac{z-1+\frac{R}{2}}{\tau}}-e^{n2\pi i \frac{-z-\frac{R}{2}}{\tau}}}{(1-e^{n2\pi i \sigma/\tau})}\Big)
\ee
It is convenient to pick the window  $z+R/2\in (0,1)$. For this case, both these terms become $1$ and the high temperature limit of $\ZZ_P(z,\sigma,\tau)$ is given by the the right hand side.
\bea
\Gamma(z+\frac{R}{2}(1+\sigma+\tau),\sigma,\tau) &\xrightarrow{\sigma,\tau\to 0^{+i}}& e^{-i\frac{\pi}{3} P_{\chi_R}(z,\sigma,\tau)}\nonumber\\
&\xrightarrow{\sigma,\tau\to 0^{+i}}& e^{-i\frac{\pi}{24}\Big((k_{RRR}-k_R)\frac{1}{\sigma\tau}+(3k_{RRR}-k_R)(\frac1\sigma+\frac1\tau)\Big)+{\cal O}(1)}.
\eea
Here $P_{\chi_R}(z,\sigma,\tau)$ is the polynomial given in equation \eqref{P-R}. In taking the high temperature limit of $P$ we have kept the anomaly terms only involving the R-symmetry.  
This reproduces the result about the high temperature limit of the elliptic gamma function given in \cite{Ardehali:2015bla}. In order to see the match, we set $x\to (z+R/2)\in (0,1)$ in equation (2.11) of \cite{Ardehali:2015bla}.
Amusingly, it is observed in \cite{Aharony:2013dha}, that in addition to taking $\sigma,\tau\to 0^{+i}$ if we take $z\to 0^{+i}$, the modular property of the elliptic gamma function reduces to an equation that expresses the index of the chiral multiplet in terms of the $3d$ partition function of the dimensionally reduced theory with all its Kaluza-Klein modes\footnote{We thank Shlomo Razamat for pointing this out to us.}. It would be interesting to investigate this further.

For gauge theories, the high temperature limit of the partition function also involves the high temperature limit of the vortex part $Z_V(z_i,\sigma,\tau)$.
 As $Z_V$ can be written explicitly in terms of theta functions of the type $\theta(z_i+n\sigma,\tau)$, its high temperature limit is obtained by taking high temperature limit of the theta function. We note  that $Z_V(z_i,\sigma,\tau)$ is also invariant under $z_i\to z_i+1$. Using this property, we restrict ourselves to computing the high temperature behavior in a suitable window of $z_i$ of width $1$.
 Using $S_{13}$ invariance of $Z_V$,
\bea
Z_V(z_i,\sigma,\tau)&=&Z_V(\frac{z_i}{\tau},\frac{\sigma}{\tau},-\frac{1}{\tau})\\
\theta(\frac{z_i+n\sigma}{\tau},-\frac{1}{\tau})&=&\exp\Big(\sum_{n=1}^\infty\frac{-e^{n2\pi i \frac{z_i+n\sigma}{\tau}}-e^{n2\pi i \frac{-1-z_i-n\sigma}{\tau}}}{(1-e^{-n2\pi i /\tau})}\Big)\xrightarrow{\tau\to 0^{+i}} \exp\Big(-e^{n2\pi i \frac{z_i}{\tau}}-e^{n2\pi i \frac{-1-z_i}{\tau}}\Big).\nonumber
\eea
It is best to pick the window $z_i\in (-1,0)$. In this case, all the theta functions involved and hence the entire vortex partition function $Z_V$ simply goes to $1$. Due to periodic symmetry $z_i\to z_i+1$, $Z_V$ becomes $1$ for all values of $z_i$.
The high temperature limit of the supersymmetric index is then obtained by the high-temperature limit of the perturbative part $Z_P$.

As we remarked early in the paper, we require the R-charges of all the fields to be integers to preserve supersymmetry. The correct superconformal R-symmetry $R'$ is obtained from this integral R-symmetry $R$ by shifting it with an abelian global symmetry $R'= R+\epsilon F$.  The coefficients of $k_{RRR}$ and $k_R$ terms in $P(z_i,\sigma,\tau)$ are unaffected by these shifts. That is, the coefficient of $k_{R'R'R'}$ is the same as the coefficient of $k_{RRR}$ and the coefficient of $k_{R'}$ is the same as the coefficient of $k_R$.  Albeit, the coefficients of all other anomaly terms such as $k_{RRF}, k_{RFF}$ and $k_F$ will be affected by this shift.  
If we are interested anomalies involving only the R-symmetry, we can simply replace $k_{RRR}$ and $k_R$ by $k_{R'R'R'}$ and $k_{R'}$  and express them in terms of  central charges $a$ and $c$. 
\be
a=9k_{R'R'R'}-3k_{R'},\qquad c=9k_{R'R'R'}-5k_{R'}.
\ee
In this way,  the high temperature limit of equation \eqref{modularYY}  reduces to the Cardy formula for the modified superconformal index given in  \cite{Kim:2019yrz, Cabo-Bizet:2019osg}.  Keeping the anomaly terms only involving the R-symmetry,
\bea
\ZZ_{S_{23}}^{\alpha\alpha}(z_i,\sigma,\tau) &\xrightarrow{\sigma,\tau\to 0^{+i}}& e^{-i\frac{\pi}{3} P(z_i,\sigma,\tau)}\nonumber\\
\ZZ_{S_{23}} (z_i,\sigma,\tau)=\sum_\alpha\ZZ_{S_{23}}^{\alpha\alpha}(z_i,\sigma,\tau) &\xrightarrow{\sigma,\tau\to 0^{+i}}& 
e^{-i\frac{\pi}{24}\Big((k_{R'R'R'}-k_{R'})\frac{1}{\sigma\tau}+(3k_{R'R'R'}-k_{R'})(\frac1\sigma+\frac1\tau)\Big)+{\cal O}(1)}\nonumber\\
&=&
e^{-i\frac{\pi}{24}\Big(\frac{3c-2a}{9}\frac{1}{\sigma\tau}+\frac{a}{3}(\frac1\sigma+\frac1\tau)\Big)+{\cal O}(1)}.
\eea
As the high-temperature behavior of $\ZZ_{S_{23}}^{\alpha\alpha}$ is independent of $\alpha$ and the index set $\alpha$ is finite, we can substitute $\ZZ_{S_{23}}^{\alpha\alpha}$ by the superconformal index $\ZZ_{S_{23}}$ as we have done in the second line.
This formula has played a crucial role in computing the entropy of supersymmetric black holes  \cite{Kim:2019yrz, Cabo-Bizet:2019osg}.
Of course, the equation  \eqref{modularYY} being detailed modular formula, may tell us more about the high energy states than just the density of states. It would be interesting to dig deeper.
 
 \subsection{Six dimensions}\label{six-d}
 In this subsection we will give a set of conjectures about supersymmetric partition functions in six dimensions.
 The index of the $(1,0)$ supersymmetric chiral multiplet  in six dimensions is known to be the double elliptic gamma function $\Gamma_2(z;\sigma,\tau,\zeta)$ \cite{Lockhart:2012vp} (see appendix \ref{double-gamma} for definition). The double elliptic gamma function is third in the hierarchy of multiple gamma functions. The first two being,
 \be
 \Gamma_0(z,\sigma)=1/\theta(z,\sigma),\qquad \Gamma_1(z,\sigma,\tau)=\Gamma(z,\sigma,\tau)
 \ee
 Interestingly, $\Gamma_0,\Gamma_1$ and $\Gamma_2$ are the supersymmetric indices of the chiral multiplets in dimension $2, 4$ and $6$ dimensions respectively. The multiple gamma functions $\Gamma_i$ enjoy interesting modular properties. They were discovered in  \cite{NARUKAWA2004247}. See appendix \ref{double-gamma} for the modular property of $\Gamma_2$. In this subsection, we will concern ourselves with the modular properties of $\Gamma_2$ and its physical origin.
 
 Although to our knowledge, this has not been established or even conjectured  in any mathematics literature, we believe $\Gamma_i$ defines a nontrivial class in $H_i({\cal G}_i,\tN/\tM)\simeq H_{i+1}({\cal G}_i,\tM)$ where ${\cal G}_i\equiv SL(i+2,{\mathbb Z})\ltimes {\mathbb Z}^{i+2}$ and that the modular property discovered in \cite{NARUKAWA2004247} is a consequence of the group relation $Y^{i+2}=(-1)^{i+1}$ where, for odd $i$, $Y$ is the element of $SL(i+2,{\mathbb Z})$ that permutes the elements cyclically and for even $i$, $Y$ is the element of $SL(i+2,{\mathbb Z})$ that permutes the elements cyclically with one element getting a minus sign. This claim, in suitable language, is well-known for $i=0$ (in this case $Y$ is the $S$-transformation of $SL(2,{\mathbb Z})$) and has been established in \cite{Felder_2000} for $i=1$. 
 
\begin{figure}[h]
    \centering
    {\includegraphics[width=0.65\textwidth]{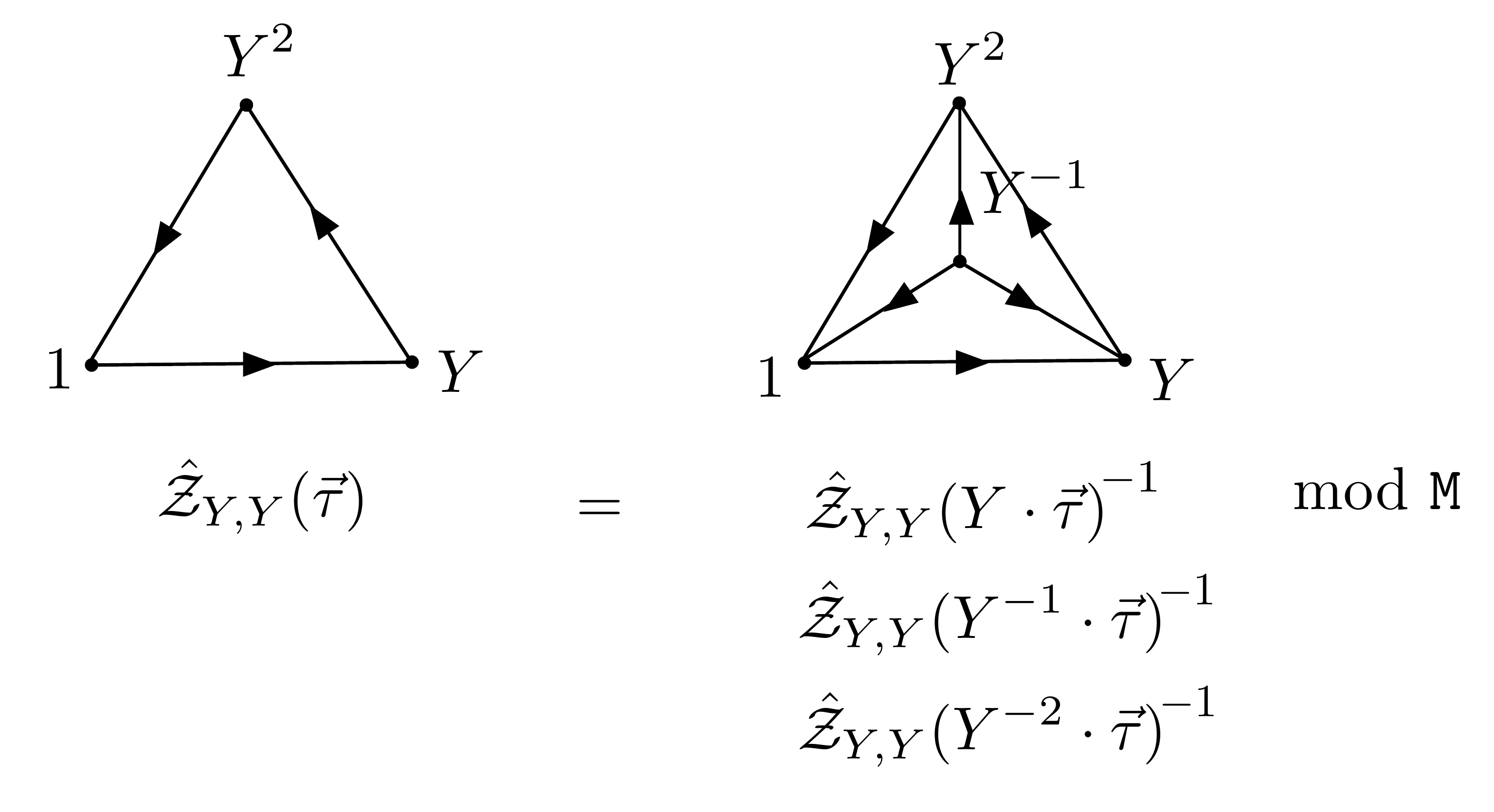} } 
    \caption{We have specialized the equation given in figure \ref{h2general} to $(g_1,g_2)=(Y,Y)$. The six-manifold corresponding to the above fibration of ${\mathbb T}^4$ i.e. given by the 2-simplex $(1,Y,Y^2)$ is precisely $S^5\times S^1$. The equation obeyed by $\hat \ZZ_{Y,Y}(z,\sigma,\tau,\zeta)$ is the same as the one obeyed by $\Gamma_2(z,\sigma,\tau,\zeta)$.}\label{h2Y}
\end{figure}

For $i=2$, we conjecture that the physical origin of this claim comes from thinking of $S^5\times S^1$ to be a ${\mathbb T}^4$ fibration over a disc, or more generally by considering the partition functions on all supersymmetry preserving six manifolds that are ${\mathbb T}^4$ fibration over a disc. We have conjectured a relation obeyed by the six dimensional partition functions on such manifolds in figure \ref{h2general}. Restricting to a special choice of group elements $g_i$, we get the equation in figure \ref{h2Y}.
Remarkably $\hat \ZZ_{Y,Y}(z,\sigma,\tau,\zeta)$ obeys the same modular equation that is obeyed by $\Gamma_2(z,\sigma,\tau,\zeta)$. 
Moreover the six-manifold corresponding to the fibration of ${\mathbb T}^4$ given by the 2-simplex $(1,Y,Y^2)$ in figure \ref{h2Y} is precisely $S^5\times S^1$. This strengthens our conjecture that supersymmetric partition functions on ${\mathbb T}^4$ fibrations over disc indeed form a non-trivial class in the cohomology $H^2({\cal G}_{6d},\tN/\tM)$ where ${\cal G}_{6d}$ consists of the large diffeomorphism group $SL(4,{\mathbb Z})$ of ${\mathbb T}^4$ and large gauge transformations. On the other hand, it also provides support for the claim that $\Gamma_2$ defines a nontrivial class in $H^2(SL(4,{\mathbb Z})\ltimes {\mathbb Z}^4,\tN/\tM)$.

In \cite{Lockhart:2012vp}, a decomposition similar to the holomorphic block decomposition has been proposed for supersymmetric partition functions on $S^5\times S^1$. It involves constructing in the index by gluing three ``holomorphic blocks" (as opposed to two in four dimensions). These blocks are related to each other by $SL(3,{\mathbb Z})$ transformations (in four dimensions the two blocks are related to each other by $SL(2,{\mathbb Z})$ transformations). However, in the same way, that $SL(3,{\mathbb Z})$ (and not $SL(2,{\mathbb Z})$) plays a fundamental cohomological role in constraining supersymmetric partition functions in four dimensions on manifolds that are ${\mathbb T}^3$ fibrations over an interval, we expect $SL(4,{\mathbb Z})$ (and not $SL(3,{\mathbb Z})$) to play a similar fundamental cohomological role in constraining supersymmetric partition functions on in six dimensions on manifolds that are ${\mathbb T}^4$ fibrations over the disc.

We plan to investigate the properties of supersymmetric partition functions in dimensions other than four in detail in the future.

\subsection*{Outlook}
In this paper, we have considered four dimensional supersymmetric theories on ${\mathbb T}^3$ fibrations on an interval. One can also consider them on ${\mathbb T}^2$ fibrations on a disc. Our discussion of group cohomology leads us to suspect that such partition functions should form a nontrivial class in $H^2(SL(2,{\mathbb Z})\ltimes {G},\tN/\tM)$ where ${G}$ is the group of large gauge transformations. Similar guesses can be made for other types of fibrations, for example, supersymmetric partition functions on ${\mathbb T}^5$ fibrations on an interval seem to form a nontrivial class in $H^1(SL(5,{\mathbb Z})\ltimes {G},\tN/\tM)$. Again, ${G}$ is the group of large gauge transformations. This class of proposals certainly require further investigations, in particular, of the question whether supersymmetry can be preserved on such fibrations. It makes for an interesting study.

This takes us to another point. All the claims and conjectures in this paper have been made for partition functions of supersymmetric theories. But, because these claims/conjectures follow simply from large diffeomorphism and large gauge transformation symmetries we expect that the versions of our statements to hold even for non-supersymmetric theories. 
For example, we believe the equation analogous to \eqref{master} and \eqref{modularY} should hold even for non-supersymmetric theories but with some crucial distinctions.
For non-supersymmetric case, the Hilbert space on ${\mathbb T}^3$ is infinite dimensional in contrast to the effectively finite dimensional Hilbert space for supersymmetric theories. The partition function $\ZZ^{\alpha\alpha}$, for supersymmetric theories,  has the meaning of partition function evaluated in a given Higgs branch vacuum and it can be computed explicitly using Higgs branch localization. This is presumably not true for non-supersymmetric theories. Also, for non-supersymmetric theories, the partition function is non-holomorphic in parameters. 
Having said all this, it would be nice to exhibit the modular properties for a  non-supersymmetric theory concretely. 
To start with, it would be interesting to do this for a free theory. 
It would also be interesting to make contact with the work of \cite{Shaghoulian_2017} where a formula has been conjectured that relates the partition function of a conformal field theory on $S^3$ at high temperature to Casimir energy on a highly lensed $S^3$.
A useful application of equations  \eqref{master} and \eqref{modularY}  for non-supersymmetric theories will open a new window into the universal properties of high energy states of higher dimensional conformal field theories.

\acknowledgments
We would like to thank  Indranil Halder for collaboration during the initial stages of this work.
We would like to thank Chris Beem, Eknath Ghate, Shiraz Minwalla, Leonardo Rastelli, Shlomo Razamat for useful discussions. We are especially grateful to Mahan Mj and Pavel Putrov for several useful conversations. We would also like to thank Shiraz Minwalla, Pavel Putrov, Leonardo Rastelli and Shlomo Razamat for extremely useful comments on the draft.
This work is supported by the Infosys Endowment for the study of the Quantum Structure of Spacetime and  by the SERB Ramanujan fellowship. We would like to acknowledge that part of this work was performed at the Aspen Center for Physics, which is supported by the National Science Foundation grant PHY-1607611. We would all also like to acknowledge our debt to the people of India for their steady support to the study of the basic sciences.

\appendix

\section{Special functions}\label{functions}
\subsection{q-theta function}\label{theta}
The q-theta function is a variant of the more famous Jacobi theta function. It is defined as
\be
\theta(z,\tau)=\exp\Big(\sum_{n=1}^\infty \frac1n \frac{-x^n-q^n/x^n}{1-q^n}\Big)\qquad {\rm where}\quad (x,q)=e^{2\pi i(z,\tau) }.
\ee
It is defined for all values except $|q|=1$. For $|q|<1$, it can be written as an infinite product,
\be
\theta(z,\tau)=\prod_{i=1}^\infty (1-x q^i)(1-q^{i+1}/x).
\ee
We will list some of its useful properties below.
\begin{itemize}
\item $\theta(z+1,\tau)=\theta(z,\tau+1)=\theta(z,\tau)$
\item $\theta(\tau-z,\tau)=\theta(z,\tau)$
\item $\theta(-z,\tau)=\theta(z+\tau)=-e^{2\pi i z}\theta(z,\tau)$
\item $\theta(z,-\tau)=-e^{2\pi i z}/\theta(z,\tau)$
\end{itemize}
Most important of all, the q-theta function has an interesting modular property,
\bea
\theta(\frac{z}{\tau},-\frac{1}{\tau})&=&e^{i\pi B(z,\tau)}\theta(z,\tau)\nonumber\\
B(z,\tau)&=&\frac{z^2}{\tau}+z(\frac{1}{\tau}-1)+\frac16(\tau+\frac1\tau)-\frac12.
\eea

\subsection{Elliptic gamma function}\label{elliptic-gamma}

The elliptic gamma function and its properties have been  discussed in great detail in \cite{Felder_2000} and references therein. It is defined as
\be
\Gamma(z,\sigma, \tau)=\exp \Big(\sum_{n=1}^\infty\frac1n \frac{x^n-\frac{p^n q^n}{x^n}}{(1-p^n)(1-q^n)}\Big),\qquad {\rm where}\quad (x,p,q)=e^{2\pi i(z,\sigma,\tau) }.
\ee
It is defined for all values of $(x,p,q)$ except for $|p|=1$ or $|q|=1$ or $x=1$.
For $|p|,|q|<1$ and $x\neq 1$, it can be written as a double infinite product,
\be
\Gamma(z,\sigma,\tau)=\prod_{i,j=0}^\infty\frac{1-p^{i+1}q^{j+1}/x}{1-x\,p^i q^j}.
\ee
We will list some of its useful properties below.
\begin{itemize}
\item $\Gamma(z,\sigma, \tau)= \Gamma(z,\tau,\sigma)$
\item $\Gamma(z+1,\sigma, \tau)=\Gamma(z,\sigma+1,\tau)=\Gamma(z,\sigma,\tau+1)= \Gamma(z,\sigma,\tau)$
\item $\Gamma(z+\sigma,\sigma, \tau)= \theta(z,\tau)\Gamma(z,\sigma,\tau)$
\item $\Gamma(z+\tau,\sigma, \tau)= \theta(z,\sigma)\Gamma(z,\sigma,\tau)$
\item $\Gamma(z,-\sigma,\tau)=\Gamma(\tau-z, \sigma, \tau)=1/\Gamma(z+\sigma,\sigma, \tau)$
\item $\Gamma(z,\sigma,-\tau)=\Gamma(\sigma-z, \sigma, \tau)=1/\Gamma(z+\tau,\sigma, \tau)$
\end{itemize}
Most important of all, the elliptic gamma function has an interesting modular property,
\bea\label{sym-modular}
\Gamma(z,\tau,\sigma)\Gamma(\frac{z}{\tau},\frac{\sigma}{\tau},\frac1\tau)\Gamma(\frac{z}{\sigma},\frac{1}{\sigma},\frac{\tau}{\sigma})&=&e^{-i\frac{\pi}{3} {Q}(z,\tau,\sigma)}\nonumber\\
{Q}(z,\tau,\sigma)
&=&\frac{z^3}{\sigma \tau}-\frac32\frac{\tau+\sigma+1}{\sigma \tau}z^2+\frac{\tau^2+\sigma^2+3\tau\sigma+3\tau+3\sigma+1}{2\tau \sigma}z\nonumber\\
&-&\frac{1}{4}(\tau+\sigma+1)(\frac1\tau+\frac1\sigma+1).
\eea
In order to make the cyclic symmetry between $(1,\sigma,\tau)$ transparent, it is better to use the homogeneous coordinates $(\omega_1,\omega_2,\omega_3)$ using $z\to \xi/\omega_1,\sigma\to \omega_2/\omega_1, \tau\to \omega_3/\omega_1$. Then,
\bea
\Gamma(\frac{\xi}{\omega_1},\frac{\omega_2}{\omega_1},\frac{\omega_3}{\omega_1})\Gamma(\frac{\xi}{\omega_2},\frac{\omega_3}{\omega_2},\frac{\omega_1}{\omega_2})\Gamma(\frac{\xi}{\omega_3},\frac{\omega_1}{\omega_3},\frac{\omega_2}{\omega_3})&=&e^{-i\frac{\pi}{3} {Q}(\frac{\xi}{\omega_1},\frac{\omega_2}{\omega_1},\frac{\omega_3}{\omega_1})}\\
\omega_1\omega_2\omega_3\,{Q}(\frac{\xi}{\omega_1},\frac{\omega_2}{\omega_1},\frac{\omega_3}{\omega_1})&=&\xi^3-\frac32 \xi^2\sum_i\omega_i+\frac{\xi}{2}(\sum_i \omega_i^2 +3\sum_{i< j} \omega_i\omega_j)\nonumber\\
&-&\frac{1}{4}(\sum_i \omega_i)(\sum_{i<j}\omega_i \omega_j)=B_{3,3}(\xi;\omega_i),\nonumber\\
&=&\xi^3-3\os \xi^2+(3\os^2-\oss)\xi+\os\oss-\os^3.\nonumber
\eea
The polynomial $B_{3,3}(\xi;\omega_i)$ is known as the Bernoulli polynomial of the third order. For convenience, we have defined $\os=\sum_i\omega_i/2$ and $\oss=\sum_i\omega_i^2/4$.

\subsection{Elliptic hypergeometric series}
The elliptic hypergeometric series is the elliptic generalization of the usual hypergeometric series. 
\be\label{ehs}
_N E_{N-1}({\vec z},{\vec \zeta};\sigma;\tau;u)=\sum_{n\geq 0}\prod_{i=1}^N \frac{\Theta
(z_i;\sigma;\tau)_n}{ \Theta
(\zeta_i;\sigma;\tau)_n} u^n,
\ee
where the theta factorial $\Theta(z;\sigma;\tau)_n$ for $n> 0$ is defined as 
\be\label{theta-fac}
\Theta(z;\sigma; \tau)_n=\prod_{j=0}^{n-1}\theta(z+j \sigma, \tau).
\ee
More generally,
\be
\Theta(z;\sigma; \tau)_n=\frac{\Gamma(z+n\sigma;\sigma,\tau)}{\Gamma(z,\sigma,\tau)}.
\ee

Alternatively, we can define the elliptic hypergeometric series using a shift operator $t_2^{(z)}:z\to z+\sigma$.
\bea\label{diff-op}
\Theta(z;\sigma; \tau)_n&=&\frac{1}{\Gamma(z,\sigma,\tau)} t_2^n \,{\Gamma(z,\sigma,\tau)}\nonumber\\
_N E_{N-1}({\vec z},{\vec \zeta};\sigma;\tau;u) &=& \prod_{i=1}^N \frac{\Gamma(\zeta_i,\sigma,\tau)}{\Gamma(z_i,\sigma,\tau)}\Big(\frac{1}{1- u \prod_{i=1}^N t_2^{(z_i)} t_2^{(\zeta_i)}}\Big) \prod_{i=1}^N \frac{\Gamma(z_i,\sigma,\tau)}{\Gamma(\zeta_i,\sigma,\tau)}.
\eea

\subsection{Double elliptic gamma function}\label{double-gamma}
Multiple elliptic gamma functions have been defined in \cite{Nishizawa_2001}. The double elliptic gamma function is,
\be
\Gamma_2(z,\sigma, \tau,\zeta)=\exp \Big(\sum_{n=1}^\infty\frac1n \frac{x^n+\frac{p^n q^n r^n}{x^n}}{(1-p^n)(1-q^n)}\Big),\qquad {\rm where}\quad (x,p,q,r)=e^{2\pi i(z,\sigma,\tau,\zeta) }.
\ee
It is defined for all values of $(x,p,q,r)$ except for $|p|=1$ or $|q|=1$ or $|r|=1$ or $x=1$.
For $|p|,|q|,|r|<1$ and $x\neq 1$, it can be written as a double infinite product,
\be
\Gamma_2(z,\sigma,\tau,\zeta)=\prod_{i,j,k=0}^\infty\frac{1}{(1-x\,p^i q^j r^k)(1-p^{i+1}q^{j+1}r^{k+1}/x)}.
\ee
Modular properties of multiple elliptic gamma functions have been discovered in \cite{NARUKAWA2004247}. This property of the double elliptic gamma function is,
\be
\Gamma_2(z,\sigma,\tau,\zeta)\Gamma_2(\frac{z}{\sigma},\frac{1}{\sigma},\frac{\tau}{\sigma},\frac{\zeta}{\sigma})\Gamma_2(\frac{z}{\tau},\frac{\sigma}{\tau},\frac{1}{\tau},\frac{\zeta}{\tau})\Gamma_2(\frac{z}{\zeta},\frac{\sigma}{\zeta},\frac{\tau}{\zeta},\frac{1}{\zeta})=e^{-i\frac{\pi}{12}B_{4,4}(\xi,\omega_i)}.
\ee
where $B_{4,4}(\xi,\omega_i),i=1,\ldots,4$ is the Bernoulli polynomial of the fourth order and $(\xi,\omega_i)$ are the homogeneous variables i.e. $(z,\sigma,\tau,\zeta)=(\xi/\omega_1,\omega_2/\omega_1,\omega_3/\omega_1,\omega_4/\omega_1)$.

\bigskip
\bibliography{cp.bib}
\end{document}